\documentclass[
  aps,
  prb,
  twoside,
  twocolumn,
  showpacs,
  floatfix,
  superscriptaddress,
  10pt,
  preprintnumbers,
  citeautoscript,
]{revtex4-1}

\usepackage{amsmath}
\usepackage{comment}
\usepackage{glossaries}
\usepackage{graphicx}
\usepackage{here}
\usepackage[utf8]{inputenc}
\usepackage{times}
\usepackage{units}
\usepackage{xcolor,colortbl}
\usepackage{hyperref}
\usepackage{titlesec}
\usepackage{booktabs}

\newcommand{\Rb}{\mathbf{R}}

\newcommand{\vb}{\mathbf{v}}
\newcommand{\eb}{\mathbf{e}}
\newcommand{\qb}{\mathbf{q}}

\newcommand{\inner}[2]{\langle#1|#2\rangle}

\hypersetup{
    unicode=false,          
    pdftoolbar=true,        
    pdfmenubar=true,        
    pdffitwindow=false,     
    pdfstartview={FitH},    
    pdfnewwindow=true,      
    colorlinks=true,        
    linkcolor=blue,         
    citecolor=blue,         
    filecolor=blue,         
    urlcolor=blue           
}

\AtBeginDocument{
\heavyrulewidth=.08em
\lightrulewidth=.05em
\cmidrulewidth=.03em
\belowrulesep=.65ex
\belowbottomsep=0pt
\aboverulesep=.4ex
\abovetopsep=0pt
\cmidrulesep=\doublerulesep
\cmidrulekern=.5em
\defaultaddspace=.5em
}

\begin{document}
\begin{abstract} 
For many materials, Raman spectra are intricately structured and provide valuable information about compositional stoichiometry and crystal quality.
Here we use density-functional theory calculations, mass approximation, and the Raman intensity weighted $\Gamma$-point density of state approach to analyze Raman scattering and vibrational modes in zincblende, wurtzite, and hexagonal BX (X = N, P, and As) structures.
The influence of crystal structure and boron isotope disorder on Raman line shapes is examined.
Our results demonstrate that long-range Coulomb interactions significantly influence the evolution of Raman spectra in cubic and wurtzite BN compounds.
With the evolution of the compositional rate from $^{11}$B to $^{10}$B, a shift toward higher frequencies, as well as the maximum broadening and asymmetry of the Raman peaks, is expected around the 1:1 ratio.
The calculated results are in excellent agreement with the available experimental data.
This study serves as a guide for understanding how crystal symmetry and isotope disorder affect phonons in BX compounds, which are relevant to quantum single-photon emitters, heat management, and crystal quality assessments.

\end{abstract}

\title{Boron Isotope Effects on Raman Scattering in Bulk BN, BP, and BAs: A Density-Functional Theory Study}

\newcommand{\wigner}{HUN-REN Wigner Research Centre for Physics, PO Box 49, H-1525 Budapest, Hungary}
\newcommand{\BME}{Department of Atomic Physics, Institute of Physics,
Budapest University of Technology and Economics, M\H{u}egyetem rkp. 3, H-1111 Budapest, Hungary}
\newcommand{\iutphysics}{Department of Physics, Isfahan University of Technology, Isfahan 84156-83111, Iran}
\newcommand{\qtf}{Quantum Technology Finland Center of Excellence, Department of Applied Physics,
Aalto University, P.O. Box 15600, FI-00076 Aalto, Finland}
\newcommand{\aaltophysics}{Department of Bioproducts and Biosystems, Aalto University, P.O. Box 16100, FI-00076 Aalto, Finland}
\newcommand{\cmmdms}{Interdisciplinary Centre for Mathematical Modelling and Department of Mathematical Sciences,
Loughborough University, Loughborough, Leicestershire LE11 3TU, United Kingdom}

\author{Nima Ghafari Cherati}
\affiliation{\wigner}
\affiliation{\BME}
\author{I. Abdolhosseini Sarsari}
\affiliation{\iutphysics}
\author{Arsalan Hashemi}
\email{arsalan.hashemi@aalto.fi}
\affiliation{\qtf}
\author{Tapio Ala-Nissila}
\email{tapio.ala-nissila@aalto.fi}
\affiliation{\qtf}
\affiliation{\cmmdms}

\maketitle
\section{Introduction}
\label{sec:introduction}

Polymorphic forms of BX (X = N, P, As) show a broad spectrum of mechanical, thermal, and optoelectronic properties\cite{Kang2019,Kang2021,Lee_apl_2024}.
These forms exist as various polytypes, including hexagonal structures (graphite-like), cubic (diamond-like), and wurtzite structures (landsdaleit-like) \cite{Mirkarimi1997MSER, Solozhenko2022_rsc}.
Hexagonal boron nitride, exhibiting a wide band gap of around 6 eV, offers a promising low-dimensional material for hosting single-photon emitter point defects \cite{Shaik2021,Ramsay2023,Pelliciari2024}.
Cubic BAs achieves an isotropic and exceptionally high thermal conductivity of 1300 W/mK, exceeding that of all common metals and semiconductors \cite{Kang2018,Li_sci2018}.
In addition, cubic BP has demonstrated highly efficient catalytic activity and durability in CO$_2$ reduction reactions \cite{Mou2019,Zhang2023_DT}.
Both composition and symmetry can be adjusted to customize the properties of BX compounds for specific applications.

Isotopic boron (B) in these compounds adds an adjustable materials parameter for application-specific engineering \cite{Vuong2018,Janzen2024AdvMat}.
More specifically, changes in isotopic mass affect (i) the lifetime and energy of the lattice dynamics\cite{Cusco_2019} subsequently reducing the thermal conductivity of the lattice\cite{Chen555,Lindsay2011}, and (ii) nuclear spins, which alter interactions with magnetic fields and the quadrupole moment, thus influencing the hyperfine interaction due to different gyromagnetic factors\cite{vuong_thesis}. Depending on the application, these effects can be either beneficial or detrimental, making it essential to determine and control the isotopic mass ratio for optimal material performance.
The complete purification of boron isotopes is challenging, which makes it essential to identify the mass defect signal. In particular, natural boron consists of about 20\% $^{10}$B and 80\% $^{11}$B isotopes.

Raman spectroscopy\cite{Yoshikawa2023,cardona1975} is a powerful, nondestructive, rapid, and widely accessible characterization tool at the nanoscale, offering detailed insight into material signatures.
This enables the detection of crucial parameters, such as crystallinity\cite{Werninghaus1997,Sachdev1997,Reich2005_PRB,Jagdish2016}, elemental composition\cite{Griffin2018,Cai2017}, strain\cite{Sachdev2003_D,solozhenko2014_JAP}, temperature\cite{Berger_prb2024,Hadjiev2014_PRB}, and defect density\cite{Cataldo2017,Gnatyuk2024opmat} in materials, which contributes to a deep understanding of their structural and functional attributes.
Because all BX materials are semiconducting\cite{Werninghaus1997}, first-order non resonant Raman spectroscopy, which covers the visible to ultraviolet range, should suffice to assess their crystal structures and quality.

Numerous experimental studies have examined the isotope effect in Raman scattering of cubic BN, BP, BAs, and hexagonal BN compounds: (i) He et al. \cite{he2021phonon} investigated B and nitrogen (N) isotope impacts on cubic BN Raman lineshapes; (ii) Zhu et al. \cite{Zhu2023_pss} studied Raman spectra of cubic BP for different isotopic ratios in detail, (iii) Rai et al. \cite{Rai_PRM2020} studied B-isotope influence on Raman of cubic BAs; (iv) both Cusc\'{o} et al. \cite{Cusco2021} and Janzen et al.\cite{Janzen2024AdvMat}, in separate studies, focused on isotope purification in hexagonal BN crystals.
Despite extensive experimental efforts and potential applications of BX, computational work on Raman spectroscopy remains limited.
Recently, Berger et al. \cite{Berger_prb2024} investigated the effect of B isotope content in \(^{11}\text{B}_x^{10}\text{B}_{1-x}\text{As}\) on Raman spectra.
However, a detailed investigation of vibrational modes and Raman signatures is still lacking.
In addition, advances in material synthesis and fabrication, along with an increasing interest in BX compounds, are expected to facilitate the development of new physical samples in the foreseeable future. Consequently, predictions of Raman spectra are highly valuable for materials that are not available.

This study aims at addressing the issue of lack of computational research on BX Raman scattering influenced by isotope effects.
We systematically examine how isotopes influence the Raman spectra of the three common allotropes: cubic, wurtzite, and hexagonal. Analyzing the Raman spectra of these structures will improve our capability to detect phase transitions and assess sample quality. We employ mass approximation and the Raman intensity weighted $\Gamma$-point density of state (RGDOS) approach in conjunction with quantum mechanical density functional theory (DFT), which have been validated in previous research\cite{Hashemi2019_PRM,Kou2020}.

\section{Methodology}
\label{sec:methodology}

The Raman effect in solids results from the inelastic scattering of light by phonons.
To first order, incident light exchanges a photon with the crystal by creating or annihilating one phonon, resulting in a gain or loss of energy in the scattered light. Raman spectroscopy measures this energy shift, enabling the determination of material's characteristic vibrational energy.

In the harmonic approximation, the vibrational modes are solutions to
\begin{equation}
 \label{eq:motion}
    M_k \omega^2 \vb(k0) = \sum_{k',l} \Phi(k'l,k0) \exp(-i\qb\cdot \Rb_l) \vb(k'0),
\end{equation}
where $\vb(kl)$ are the eigenvectors for the displacement of atom $k$ with mass $M_k$
located in cell $l$ specified by the lattice vector $\Rb_l$ at wave vector $\bf{q}$.
For $N$ number of atoms in the supercell (SC), the $\Phi_{\alpha\beta}(k'l',kl)$ element of the matrix of the force constant (FC) $3N \times 3N$ is defined by
the change of the surface of the potential energy when the $k^{\rm th}$ atom in the $l^{\rm th}$ primitive unit cell (PUC) and
the $k'^{\rm th}$ atom in the $l'^{\rm th}$ PUC are displaced in the Cartesian directions $\alpha$ and $\beta$, respectively.

In the case of isotope disorder, only the atomic masses change in Eq. (\ref{eq:motion}), while the nature of the chemical bonds, i.e., FCs, remains intact.
Thus, to compute the FC matrix for a mixture of isotopes, the masses must be distributed randomly through a large SC.
Together, we can expand the PUC FC to the SC FC and evaluate the vibrational modes of the isotopically mixed structures.

If we consider an SC, the $\Gamma$ point hosts several modes from the folding of PUC phonon bands.
By projecting the folded modes to the Raman-active vibrational modes of PUC, one can weigh the population of Raman-active modes at the SC $\Gamma$ point as follows:
\begin{equation}
    w_{ij} = \inner{\vb^{{\rm{PC}},i}}{\vb^{{\rm{SC}},j}}  =
    \sum_{\alpha,k,l} v_{\alpha}^{{\rm{PUC}},i}(k0) v_{\alpha}^{{\rm{SC}},j}(kl).
    \label{eq:wij}
\end{equation}
The Raman tensor of the SC mode is derived by multiplying the PC mode projection with the corresponding Raman tensors from the pristine system.

\begin{equation}
R^{{\rm{SC}},j} = \sum_{i} w_{ij} R^{{\rm{PUC}},i},
\label{eq:rscj}
\end{equation}
where the sum goes over PUC modes $i$ and only Raman-active modes contribute.
In the case of nonresonant first-order Raman scattering, the corresponding tensor $R^{{\rm{PC}},i}$ is obtained from the change of polarizability $\chi$ with respect to the phonon eigenvectors $\vb$, and in the DFT calculations, it can be evaluated by using the macroscopic dielectric constant
$\varepsilon_{\rm mac}$ as
\begin{equation}\label{eq:Rpol}
R \approx \frac{\partial \chi}{\partial \vb} 
  = \frac{\partial \varepsilon_{\rm mac}}{\partial \vb}.
\end{equation}
Finally, the Raman intensity of the SC mode $j$ is obtained as
\begin{align}
I^{{\rm{SC}},j} &\approx \rvert \eb_s \cdot R^{{\rm{SC}},j} \cdot \eb_i \rvert^2,  \label{eq:RGDOS}
\end{align}
where $\eb_i$ and $\eb_s$ are the polarization vectors of incident and scattered light, respectively.
The output intensity from the above procedure is that obtained in the RGDOS method.

\section{Computational details}
\label{sec:computational_details}

DFT calculations were conducted using the VASP code\cite{kres1} with projector-augmented-wave\cite{PAW1994}(PAW) pseudopotentials. The exchange-correlation contributions for zincblende and wurtzite structures utilized the Perdew-Burke-Ernzerhof revised for solids\cite{PBEsol2008} (PBEsol).
Despite the hexagonal structure, the Perdew-Burke-Ernzerhof\cite{PBE1996PRL} (PBE) functional, accompanied by the dispersion interaction (D3) is used.
The dispersion interaction is modeled by Grimme\cite{Grimme2010_D3} (GD3), Becke-Johnson\cite{Grimme2011_JCC} (BJD3) and Tkatchenko-Scheffler\cite{Tkatchenkoprb_2009} (TSD3) for the layered hexagonal bulk structures.
Plane waves with kinetic energy below 520 eV are included in the basis set.
The structural relaxation was kept running until the total energy difference and ionic forces are converged to less than 10$^{-6}$ eV and 10$^{-3}$ eV/\AA, respectively.
During structural relaxation, the Brillouin zone (BZ) was defined using 15$\times$15$\times$15, 9$\times$9$\times$5, and 10$\times$10$\times$3 grids for cubic, wurtzite, and hexagonal structures, respectively. Lattice vibration modes were assessed using the Phonopy package\cite{TOGO2015} with the finite difference method. The zincblende, wurtzite, and hexagonal structures were generated with 5$\times$5$\times$5, 5$\times$5$\times$3, and 5$\times$5$\times$2 PUC repetitions, respectively.
The long-range Coulomb interaction of the macroscopic electric field near the $\Gamma$-point was applied to all the phonon calculations, as described in Ref. \onlinecite{Pick1970_PRB}.
We statistically sampled the isotope mass distribution by generating 20 random configurations for each composition, with the spectrum for each concentration being the average of these 20 spectra.
Gaussian fitting with a broadening of 5 cm$^{-1}$ was used to obtain the lineshape of the RGDOS signals. The calculated lineshapes represent unpolarized Raman spectra.

\section{Results}
\label{sec:results}

\subsection{Structural Properties}
\label{sec:str}
\begin{figure}[!htbp]
 \centering
  \includegraphics[width=0.85\linewidth]{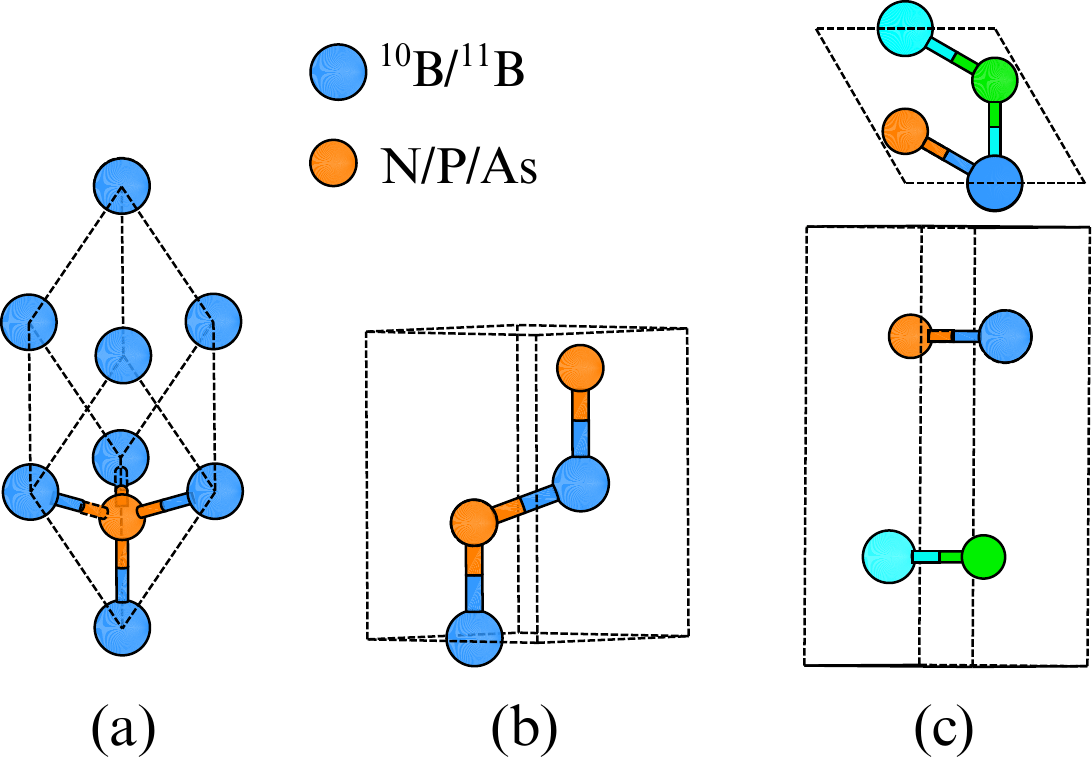}
   \caption{Primitive unit cells of various BX polytypes: (a) cubic, (b) wurtzite, and (c) hexagonal, with the hexagonal structure shown from both side and top views. Blue circles represent B atoms, and orange circles represent X atoms (where X = N, P, As). In (c), X and B in the second layer are depicted as green and cyan circles, respectively.}
    \label{fig:pucs}
\end{figure}
%

In the zincblende structure of BX (where X = N, P, or As), boron and X atoms are $sp^3$-hybridized, forming a tetrahedral network. The lattice consists of two interpenetrating face-centered cubic lattices: one for boron (B) and the other for X, resulting in a primitive unit cell with one boron and one X atom, as illustrated in Fig. \ref{fig:pucs} (a). The calculated lattice constants for BN, BP, and BAs are 2.55, 3.20, and 3.38 \AA, respectively, with experimental values of 2.55 \AA \cite{Jagdish2016}, 3.22 \AA \cite{Popper1957nat,Kang2017_nl}, and 3.38 \AA \cite{Lee_apl_2024}. These values were converted to the primitive translation vector length using $\sqrt{2}a/2$. The zincblende structure, henceforth labeled with the prefix $c$, crystallizes in cubic symmetry with a $T_{d}$ point group, indicating high crystallographic symmetry.

The wurtzite structure, denoted by $w$, has a hexagonal close-packed symmetry with the point group $C_{6v}$. It features an ABAB stacking sequence along the $c$-axis ([0001]) with vertically stacked hexagonal layers of atoms. This structure exhibits lower symmetry than zincblende $c-$BX, which slightly influences the tetrahedral bond angles and lengths, though it remains $sp^3$-hybridized. The PUC consists of four atoms, as shown in Figure \ref{fig:pucs} (b). The lattice parameters for BN, BP, and BAs are 2.54 \AA~(1.65), 3.18 \AA~(1.66), and 3.36 \AA~(1.66), respectively.

\begin{table}[!htbp]
\centering
\caption{The energy difference per unit cell, $\Delta E$, and interlayer distance, $d$, were calculated using the TSD3 method to describe van der Waals interactions (vdW).}
\label{tab:info}
\begin{tabular}{lccccccccc}
    \toprule
    & \multicolumn{2}{c}{BN}& \multicolumn{2}{c}{BP}& \multicolumn{2}{c}{BAs}\\
    & $\Delta E$ (meV) & $d$ (\AA) & $\Delta E$ (meV) & $d$ (\AA) & $\Delta E$ (meV) & $d$ (\AA) \\
    \midrule
 AA    & 63   & 3.44 &  158 & 3.97 & 176 &  4.02 \\
 AA$'$ & 13   & 3.32 &  122 & 3.83 & 138 &  3.95  \\
 AB    & 5    & 3.31 &   16 & 3.61 & 19  &  3.70 \\
 AB$'$ & 0    & 3.27 &  0   & 3.61 & 0   &  3.69 \\
 A$'$B & 52   & 3.45 &  173 & 3.68 & 81  &  3.75 \\
 \midrule
\end{tabular}
\end{table}
We calculated the total energy of various stacking orders to investigate the layered structures of hexagonal BX ($h-$BX).
In the bulk phase, a PUC consists of two layers that can be arranged in five configurations: 
1. AA (B$^{1}$ over B$^{2}$ and X$^{1}$ over X$^{2}$);
2. AB (B$^{1}$ over X$^{2}$, X$^{1}$ (B$^{2}$) at the center of the hexagon);
3. AB (B$^{1}$ over B$^{2}$, X$^{1}$ (X$^{2}$) at the center of the hexagon);
4. BA (X$^{1}$ over X$^{2}$, B$^{1}$ (B$^{2}$) at the center of the hexagon);
5. AA (B$^{1}$ over X$^{2}$ and X$^{1}$ over B$^{2}$).
Superscripts 1 and 2 denote the two layers. A structure can be transformed into another by rotating one plane around the $z$-axis and adjusting its position if necessary.

For bulk $h-$BX compounds, the AB configuration is the most stable (see Table \ref{tab:info}). In $h-$BN, the energy differences between AB and AA configurations are minimal, at 5 meV and 13 meV, respectively. Experimental studies have observed the coexistence of various stacking orders, including AB and AA, in few-layer hBN \cite{Warner2010_acs_nano,Kenji2021sci}. We employed GD3 and BJD3 methods to assess the influence of van der Waals interactions on stacking order predictions, yielding nearly identical results (details in Table S1). The AB configuration remains the most stable, followed closely by AB and AA, although lattice relaxation indicates instability in the AB configuration, which stabilizes with the TS method. For $h-$BP and $h-$BAs, AB stacking is 16 meV and 19 meV more stable than AA, respectively. Our analysis focuses on bulk structures; results may differ for 2D structures. This research emphasizes high-energy Raman-active modes and the effects of stacking order on shear (low-frequency) modes \cite{Luo2015sci}. Interested readers can refer to Refs. \onlinecite{Gilbert_2019,Constantinescu2013-PRL,MaromPRL2010} for a more in-depth discussion. To maintain consistency among the compounds, we will proceed with the AB configuration.

For BN, BP, and BAs, the values of $a~(c/a)$ are 2.50 (2.61), 3.20 (2.26), and 3.38 (2.19), respectively. The computed data for hBN closely matches the experimental findings\cite{Henck2017_PRB}, which reported $a = 2.5$ and $c/a = 2.64$. The interlayer spacings for BN, BP, and BAs are 3.27, 3.61, and 3.69, corresponding to the radii of the X atoms. The $h-$BX structures exhibit hexagonal symmetry and belong to the space group $D_{6h}$.

\subsection{Vibrational Properties}
\label{subsec:vibproperties}

\begin{figure*}[!htbp]
    \centering
 \includegraphics[width=0.98\linewidth]{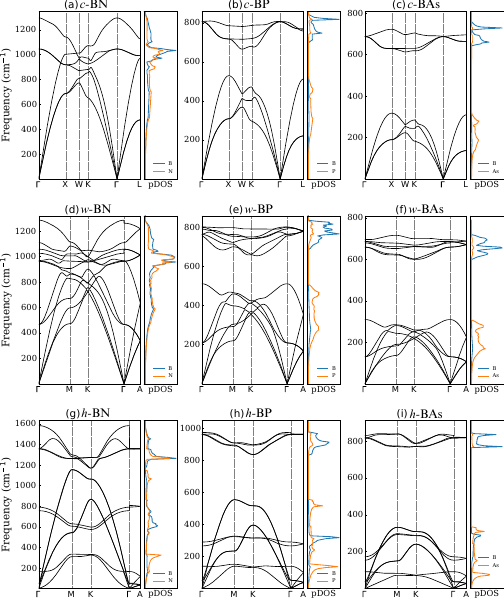}
    \caption{Phonon dispersion curves and phonon density of states (PDOS) for cubic, wurtzite, and hexagonal $^{11}$BX (where X = N, P, As) along high symmetry directions in the first Brillouin zone.
    }
    \label{fig:phonon_curves}
\end{figure*}
%

Two atoms in the PUC of $c-$BX yield six vibrational modes: three acoustic modes with zero frequency at the $\Gamma$ point and three optical modes that are both infrared (IR) and Raman-active. The irreducible representation is $\Gamma_{\rm{vib}} = 2T_{2}$ (see TABLE S2). The $T_{2}$ modes are triple-degenerate, with the B and X atoms vibrating oppositely along one dimension in the unit cell (cf. Figure S1).

Atomic contributions across the Brillouin zone are affected by mass differences among constituent atoms. The phonon density of states (pDOS) in $c-$BN shows nearly equal contributions from B and N atoms (Fig. \ref{fig:phonon_curves} (a)). Due to the larger mass difference between B and P/As, the pDOS bands are well separated (see Fig. \ref{fig:phonon_curves} (b-c)). The phonon gap is widest in $c-$BAs and absent in $c-$BNs. P and As atoms dominate in the low-energy range, while B atoms are more active in the high-energy range.

The phonon dispersion curves of $c-$BN exhibit a distinction between longitudinal optical (LO) and transverse optical (TO) vibrational modes, a phenomenon linked to the bond length-dependent ionicity in BN compounds \cite{Sanjurjo83_PRB}. The doubly degenerate TO mode occurs at 1045 cm$^{-1}$, while the single LO mode is at 1294 cm$^{-1}$. In comparison, the optical vibrational modes of $c-$BP and $c-$BAs, found at 806 cm$^{-1}$ and 686 cm$^{-1}$, respectively, are nearly triply degenerate, with a minimal difference of $\omega_{\rm{LO}} - \omega_{\rm{TO}}$ = 1 cm$^{-1}$. The vibrational modes reach up to 1300 cm$^{-1}$ for $^{11}$BN, 800 cm$^{-1}$ for $^{11}$BP, and 730 cm$^{-1}$ for $^{11}$BAs, reflecting an inverse relationship with the compounds’ reduced masses: $\mu_{\rm{BN}} = 6.2$, $\mu_{\rm{BP}} = 8.1$, $\mu_{\rm{BAs}} = 9.6$.

%
The four atoms in the PUC of $w-$BX produce three acoustic and nine optical vibrational modes, expressed as $\Gamma_{\rm{vib}} = 2E_1 + 2E_2 + 2A_1 + 2B_1$. Group theory indicates that the $A_1$ and $E_1$ modes at the BZ center are acoustic, while the two $B_1$ modes are silent. The remaining $A_1$ mode is Raman-active, and both $E_1$ and $2E_2$ modes are active in IR and Raman (see Figure S2 and TABLE S3).

For simplicity, $2E_2$ is represented as $E_{2}^{1} + E_{2}^{2}$, where $E_{2}^{1}$ denotes the lower energy mode and $E_{2}^{2}$ the higher energy mode. In the double-degenerate $E_{2}^{1}$ mode, atoms vibrate within hexagonal planes: atoms in the {\bf{BA}} layers of the A{\bf{BA}}B stacking sequence move in one direction, while others move oppositely. The frequencies for this mode are 467 cm$^{-1}$ in $^{11}$BN, 209 cm$^{-1}$ in $^{11}$BP, and 130 cm$^{-1}$ in $^{11}$BAs (Fig. \ref{fig:phonon_curves} (d-f)). In the double-degenerate $E_{2}^{2}$ mode, B and X atoms in the {\bf{AB}} layers of the {\bf{AB}}AB move together, while the other B and X atoms move in the opposite direction, all within the hexagonal plane. This mode occurs at frequencies of 963 cm$^{-1}$ in $^{11}$BN, 765 cm$^{-1}$ in $^{11}$BP, and 660 cm$^{-1}$ in $^{11}$BAs crystals.

In the out-of-plane $A_1$ mode, nearest neighbors displace oppositely, with boron atoms moving in one direction and X atoms in the opposite. In the double-degenerate $E_1$ mode, atoms vibrate within the BX hexagonal plane, again with boron and X atoms moving in opposite directions. For $w-^{11}$BN, the $A_1$ mode is observed at 1026 cm$^{-1}$ (Fig. \ref{fig:phonon_curves} (g)), while the $E_1$ mode is split into TO and LO modes at 1058 and 1286 cm$^{-1}$ ($\omega_{\rm{LO}} - \omega_{\rm{TO}}$ = 228 cm$^{-1}$). The degenerate $E_1$ mode is at 962 cm$^{-1}$. For $^{11}$BP, the Raman-active modes are the $A_1$ mode at 792 cm$^{-1}$ and the $E_1$ mode at 803 cm$^{-1}$ (Fig. \ref{fig:phonon_curves} (h)). The Raman-active $A_1$ and $E_1$ modes of $^{11}$BAs are at 680 cm$^{-1}$ and 686 cm$^{-1}$, respectively (Fig. \ref{fig:phonon_curves} (i)).
%
The PUC of $h-$BX comprises four atoms in two layers, resulting in nine optical vibrational modes categorized as $2E_{1u} + 2A_{2u} + 2E_{2g} + 2B_{1g}$ (see TABLE S4). The $E_{1u}$ and $A_{2u}$ modes are acoustic with zero vibrational frequency at the $\Gamma$ point. There are three infrared-active modes: $E_{1u}$ and $A_{2u}$. The $B_{1g}$ modes are silent, while the $2E_{2g}$ modes are Raman-active and doubly degenerate.

Our calculations show no negative energies (soft modes), highlighting the stability of the AB structures. Additionally, as anticipated for lamellar materials, there are two acoustic modes with linear $q$-dependence for the transverse and longitudinal modes in the plane, and one mode with a parabolic energy dependence that describes the out-of-plane flexural response to stress when the layers bend.

The Raman-active modes \(E_{2g}\) are shown in Fig. S2. The low-frequency modes, \(E_{2g}^{1}\) and \(E_{2g}^{2}\), are shear modes where, in \(h-\)BX, two planes vibrate in opposite directions while atoms within each plane move uniformly. These modes occur at 53, 51, and 37 cm\(^{-1}\) for \(^{11}\)BN, \(^{11}\)BP, and \(^{11}\)BAs, respectively. In contrast, the high-frequency modes \(E_{2g}^{3}\) and \(E_{2g}^{4}\) in \(h-\)BX involve B and X atoms vibrating in-plane in opposite directions, with atoms in one plane moving oppositely to those in the other. Their frequencies are 1367, 965, and 818 cm\(^{-1}\) for \(^{11}\)BN, \(^{11}\)BP, and \(^{11}\)BAs, respectively. 

In \(^{11}\)BN, the IR-active \(E_{1u}\) mode separates into LO and TO modes with a frequency difference of 231 cm\(^{-1}\), aligning well with experimental results \cite{Serrano2007}.
The LO-TO splitting in \(^{11}\)BP is approximately 14 cm\(^{-1}\), whereas it is negligible in \(^{11}\)BAs.

\subsection{Raman scattering}
\label{subsec:raman}
\begin{figure*}
 \centering
  \includegraphics[width=0.95\linewidth]{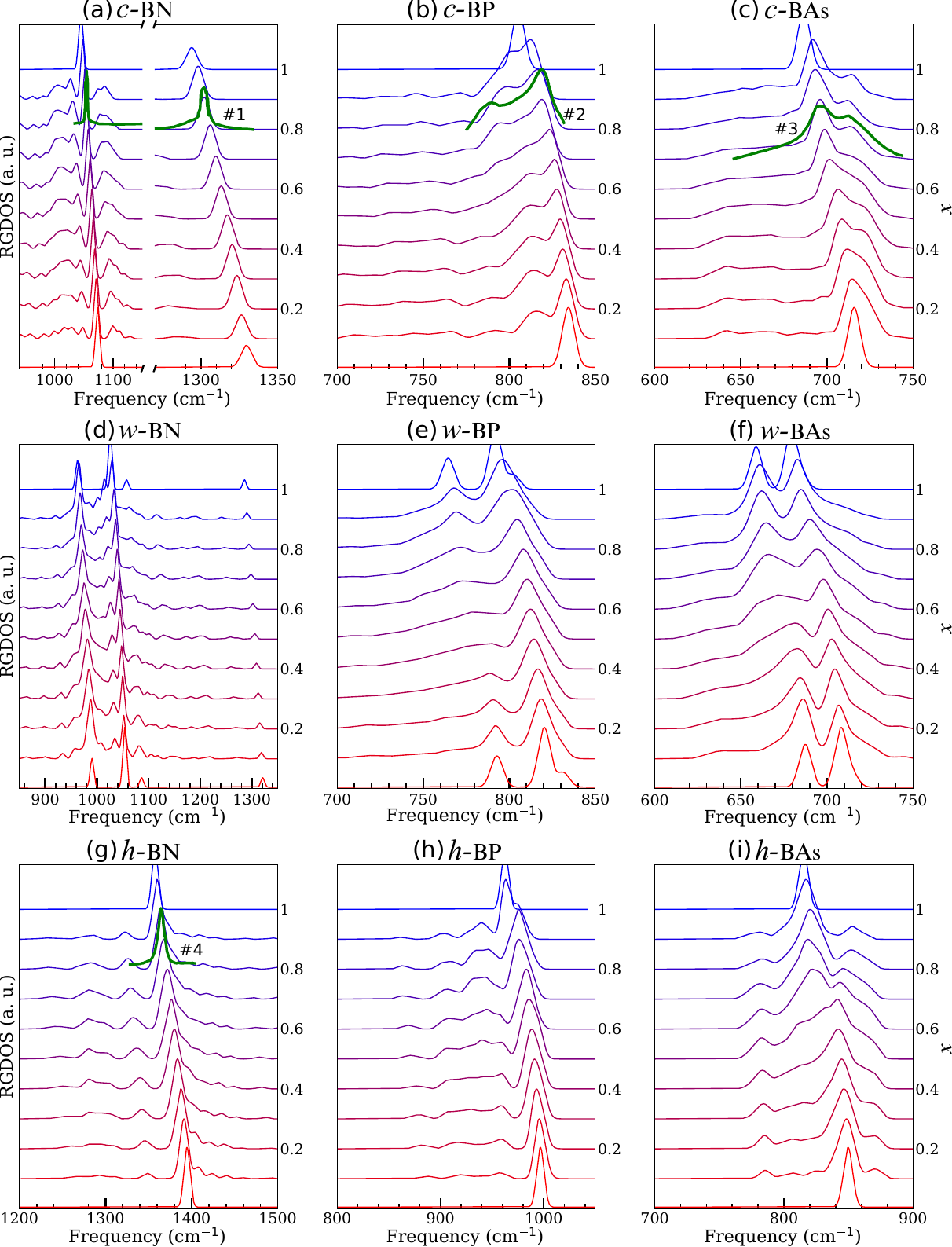}
   \caption{The RGDOS spectra of $^{11}$B$_{x}$ $^{10}$B$_{(1-x)}$X compounds (X = N, P, As) are presented for various structures. The variable $x$ indicates the concentration of $^{10}$B, ranging from 0.0 to 1.0 in increments of 0.1. The green lines labeled \#1, \#2, \#3, and \#4 correspond to experimental data from Refs. \citenum{he2021phonon}, \citenum{Zhu2023_pss}, \citenum{Rai_PRM2020}, and \citenum{Janzen2024AdvMat}, respectively. We note that datasets \#3 and \#4 were shifted down by 8 and 5 cm\(^{-1}\), respectively, to align with the computational results.}
    \label{fig:cubic}
     \end{figure*}

The naturally isotopic $c-$BN exhibits two primary peaks corresponding to the TO and LO phonons at 1051 and 1302 cm\(^{-1}\), respectively. Our findings align closely with the experimental frequencies of 1055-1058 cm\(^{-1}\) (TO) and 1304-1307 cm\(^{-1}\) (LO)~\cite{Werninghaus1997,Sachdev2003_D,Sachdev1997,he2021phonon}.
Taking into account the isotopic mass effect, several additional peaks appear around the TO mode that aren't found in the experimental data. To explore this discrepancy, we conducted direct Raman calculations for a $3\times 3 \times 3$ $c-^{\rm nat}$BN, where the superscript indicates a naturally occurring mixture of isotopes 10 and 11 of B as mentioned in the Introduction (see Figure S4). Our findings indicate that while RGDOS overestimates the intensity of these additional peaks, they are present and show a similar intensity trend.

Defect-induced peaks can be explained by the phonon confinement model (PCM) \cite{Mignuzzi2015,Richter1981}. In finite crystals, vibrational modes are spatially confined due to isotope disorder, leading to momentum uncertainty that enables vibrational modes with \({q} \ne 0\) to contribute to the Raman spectrum.
The lower-frequency feature likely stems from the downward bending of the TO band, while the upper peak results from the folding of the LO vibrational mode at the K point. Reference \onlinecite{Haque2021_ACS} attributes a similar characteristic to LO at the L point, which may be misrepresented, as it only influences the lower-frequency peaks. In transitioning from \( ^{11}\text{B} \) to \( ^{10}\text{B} \), the TO and LO peaks in $c-$BN shift upward by approximately 29 and 35 cm\(^{-1}\), respectively.

The RGDOS of $c-$BP with natural B-isotopes exhibits two peaks: a weaker one at 799 cm$^{-1}$ and a stronger one at 817 cm$^{-1}$. The calculated line shapes align closely with experimental spectra \cite{Zhu2023_pss}. The observed frequencies match experimental values of 794-799 and 823-829 cm$^{-1}$ \cite{Chen555,Kang2017_nl,Padavala2018,Zheng2018,Zhu2023_pss}. Notably, the Raman peak shifts down by approximately 29 cm$^{-1}$ from $^{11}$B to $^{10}$B, where only a single peak is detected.
This aligns with the observed frequency blue-shift of 24 cm$^{-1}$ \cite{Zhu2023_pss}. The frequency and broadening of the double-peak correlate with the isotope ratio, with broadening peaking at a 1:1 ratio.

The Raman-active LO(TO) modes in $c-$BAs show that both $^{11}$BAs and $^{10}$BAs exhibit a single symmetric peak, which upshifts by approximately 29 cm$^{-1}$, aligning with the measured values of 28 cm$^{-1}$ \cite{Hadjiev2014_PRB, Sun-material2019}. A two-mode behavior arises when $^{11}$B partially substitutes $^{10}$B. In the naturally occurring B isotope mixture, the $^{11}$B-like phonon appears at a lower frequency of 687 cm$^{-1}$, while the $^{10}$B-like phonon peaks at 715 cm$^{-1}$. Experimental peaks are observed between 700-704 cm$^{-1}$ and 720-723 cm$^{-1}$ \cite{Kang2018, Sun-material2019, Hadjiev2014_PRB, Rai_PRM2020, Lee_apl_2024}.
Our results show similar broadening to the experiments\cite{Rai_PRM2020}. The peaks related to isotope disorder broaden on the higher energy side, peaking at a concentration of $x=0.5$, which aligns with the experimental data, especially regarding the peak shapes \cite{Hadjiev2014_PRB, Rai_PRM2020}.

%
The compound $w-$BN displays four distinct peaks at 962, 1025, 1057, and 1285 cm$^{-1}$ for the $^{11}$B-rich sample, as predicted by vibrational modes analysis. Substituting $^{11}$B with $^{10}$B causes upward shifts of 28, 28, 29, and 35 cm$^{-1}$, respectively. Mass defects lead to broader lattice mode peaks and the emergence of new features across a wide frequency range.

In pristine $w-$BP, a symmetric peak appears with an accompanying shoulder from the doubly degenerate $E_2^2$ mode. In the $^{11}$B-rich case, these features are centered at 764, 791, and 802 cm$^{-1}$, while for $^{10}$BP, they are at 792, 820, and 831 cm$^{-1}$. In $w-$BP mixtures, broadening is maximized at equal shares of both isotopes ($x = 0.5$). Deviating from this ratio sharpens and localizes the flat plateau.

The RGDOSs of $w-$BAs with pure B mass exhibit two peaks at 687 and 710 cm$^{-1}$ for $^{10}$B-rich samples, which are approximately 29 and 28 cm$^{-1}$ higher than those for $^{11}$B-rich samples. The higher frequency peak consists of $A_1$ and $E_2^2$ modes, differing by 6 cm$^{-1}$. As observed in other systems, the broadening of the two-mode peak is greatest at a 1:1 isotope mass ratio.

%
We next focus on the high-frequency Raman mode of $h-$BX. Readers interested in the low-frequency \( E_{2g} \) mode can refer to Ref. \onlinecite{Janzen2024AdvMat} for $h-$BN. Isotope substitution causes phonon frequencies to shift upward from \(^{11}\text{B}\) to \(^{10}\text{B}\), with shifts of 37, 36, and 35 cm\(^{-1}\) for BN, BP, and BAs, respectively. Our calculations align with the experimental data \cite{Vuong2018,Yuan2019,Chng2020,Cusco2021,Giles2018-Natmat,Haque2021_ACS}.

In $h-$BN, the isotope effect broadens the main peak's bandwidth as compared to pure compounds. Isotope mass fluctuations result in two new peaks at lower energy than the main peak across all compositions. Similar peaks are noted in Ref. \onlinecite{Cataldo2017}. Additionally, evidence for the minor peaks at lower frequencies can be found in Jishad's experiment \cite{Jagdish2016}, which has not been widely recognized.

Isotope disorder manifests as a broad plateau in the lower frequency range of the main peak in $h-$BP, peaking at $x$ = 0.5. In $h-$BAs, mass defects significantly alter the peak shape, as illustrated in Fig. \ref{fig:cubic}. The main peak broadens with shoulders on both sides, with the widest lineshape occurring at the highest disorder. The defect-induced peak remains at a consistent energy across varying rates.

We analyzed the RGDOS spectrum of the most disordered composition ($x$ = 0.5) to identify element-wise contributions.
Each element's contribution in the SC is quantified by projecting the phonon eigenvector $j$ onto the atoms.
Thus, we can write
\begin{equation}
\left| \vb^{{\rm{SC}},j} \right|^{2} = \left| \vb^{{\rm{SC}},j}_{^{10}\rm{B}} \right|^{2} + \left| \vb^{{\rm{SC}},j}_{^{11}\rm{B}} \right|^{2} + \left| \vb^{{\rm{SC}},j}_{\rm{X}} \right|^{2},
\label{eq:partialw}
\end{equation}
where $\vb^{{\rm{SC}},j}$ is normalized to unity.
For instance, the contribution from $^{10}$B atoms can be calculated as
\begin{equation}
    \left| \vb^{{\rm{SC}},j}_{^{10}\rm{B}} \right|^{2} = \sum_{n} \left| \phi_{n,x} \right|^{2} +  \left| \phi_{n,y} \right|^{2} +  \left| \phi_{n,z} \right|^{2}.
\end{equation}
Here, $\phi_{n,\alpha}$ denotes the eigenvector of atom $n$ in the $\alpha$-direction and the sum goes over all $^{10}$B atoms in the system.

By considering Eq. (\ref{eq:rscj}) and integrating it with Eq. (\ref{eq:partialw}), we can write
\begin{equation}
R^{{\rm{SC}},j}_{c} = R^{{\rm{SC}},j} \times \left| \vb^{{\rm{SC}},j}_{c} \right|^{2},
\label{eq:rscj2}
\end{equation}
where $R^{{\rm{SC}},j}_{c}$ is defined as a compositional Raman tensor.

Partial RGDOSs for all compounds are depicted in Figure S5. (i) In BN compounds, the contributions from \(^{10}\)B, \(^{11}\)B, and N are nearly equal, with N being slightly dominant in the main peaks of $w-$BN. (ii) In BP compounds, \(^{10}\)B shapes the overall spectrum, particularly in $c-$BP and $w-$BP, while phosphorus peaks align with the main peak, and \(^{11}\)B adds a shoulder in the lower frequency range. (iii) In BAs compounds, As shows minimal Raman activity, with B isotopes shaping the spectrum in each allotrope; notably, \(^{11}\)B is more active than \(^{10}\)B at lower frequencies.

\section{SUMMARY AND CONCLUSIONS}
\label{sec:conclusions}
In this work we have systematically investigated the structural and vibrational properties, including Raman signatures, of BX (X = N, P, As) compounds, with a specific focus on isotopic effects in zincblende, wurtzite, and hexagonal structures.
Our main findings are as follows:
(i) bulk $h$-BX stabilizes in an AB$'$ stacking order;
(ii) phonon frequencies are inversely related to the reduced mass of the BX components, while the phonon gap is directly correlated with it;
(iii) The LO-TO frequency separation in BN compounds ($\sim230$ cm\(^{-1}\)) significantly changes the Raman lineshapes compared to BP or BAs, and
(iv) In BN compounds, B isotopes and N element have nearly equal contributions. In BP, \(^{10}\)B dominates, while in BAs, B isotopes primarily influence the spectral lines with minimal impact from As.
We also find that the Raman signatures are distinct across all the systems studied, enabling the detection of structural quality.
Although the BN compounds show some additional peaks as compared to predictions from polarizability-based Raman scattering calculations, the peak shapes and frequencies for $c-$BP and $c-$BAs align well with experimental data.
These findings are significant for detecting mass defects and structural crystallinity in semiconducting materials, in particular for designing quantum single-photon emitters, as well as for heat management and optoelectronic devices.
Overall, this work offers new insights into how isotope and structural symmetry affect Raman spectra, laying the groundwork for future research in semiconducting materials. Additionally, it enhances our ability to rapidly interpret experimental results and aids in characterizing polymorphic compounds.

\section{Acknowledgements}
\label{sec:acknowledgements}
We are grateful to  CSC--IT Center for Science Ltd. and
Aalto Science-IT project for generous grants of computer time.
T.A-N. and A.H. have been supported by the Academy of Finland through its QTF Center of Excellence program (Project No. 312298).

\setcounter{equation}{0}
\setcounter{table}{0}
\setcounter{figure}{0}

\renewcommand{\theequation}{S\arabic{equation}}
\renewcommand{\thepage}{S\arabic{page}}
\let\oldthetable\thetable
\renewcommand{\thetable}{S\oldthetable}
\renewcommand{\thefigure}{S\arabic{figure}}

\newpage

\section{Supplemental Material}
Supplemental materials, including additional figures, and methodological appendices, are provided in the following sub-sections.
\subsection{Different Stacking-Orders of Hexagonal BX}
\label{sec:stacking}
\begin{figure}[ht!]
    \centering
 \includegraphics[width=0.9\linewidth]{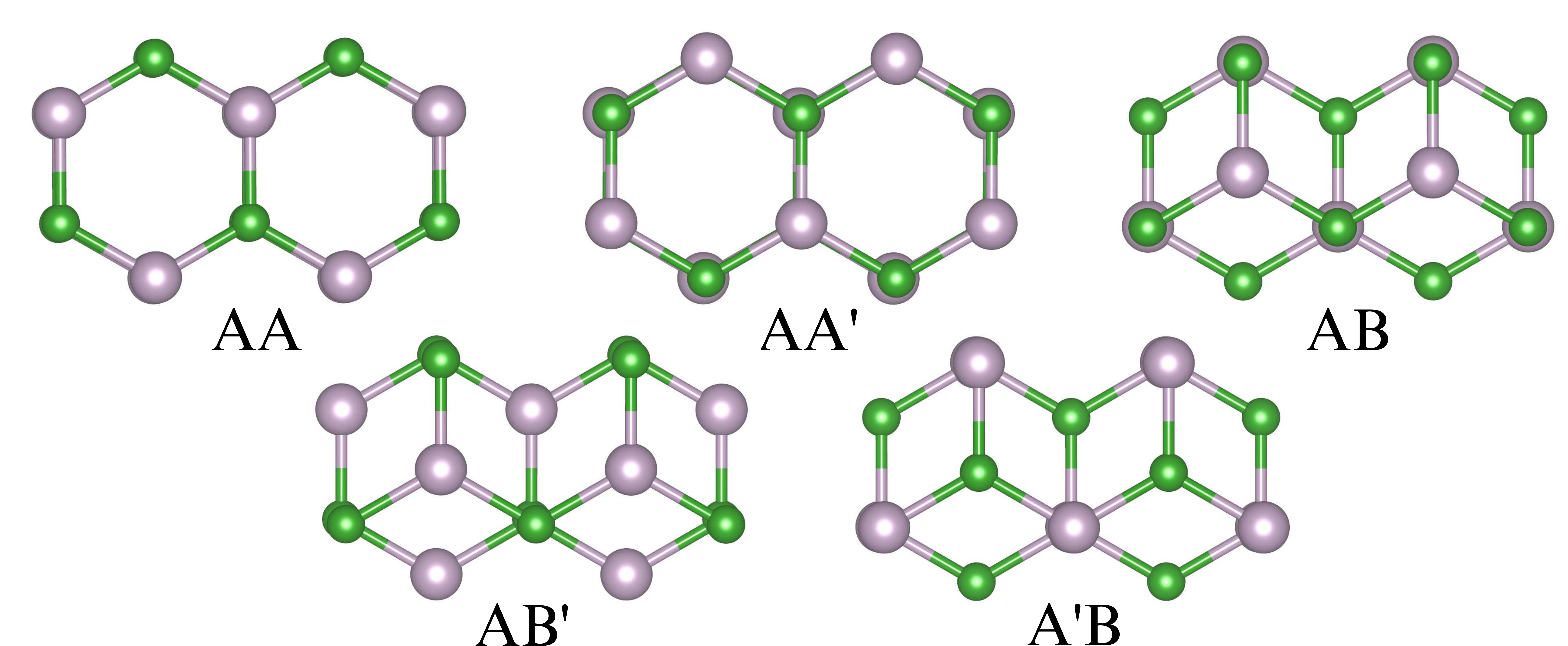}
    \caption{Distict stacking-order for $h-$BX compounds, where X = N, P, and As. Green and purple balls represent B and X elements.}
    \label{fig:stacking}
\end{figure}
The $h-$BX (X = N, P, and As) exhibits five stacking arrangements, as shown in Figure \ref{fig:stacking}.
Structural stability was evaluated for different dispersion corrections, namely D3 Grimme (GD3) \cite{Grimme2010_D3}, Becke-Johnson (BJD3) \cite{Grimme2011_JCC}, and Tkatchenko-Scheffler (TS) \cite{Tkatchenkoprb_2009} as implemented in VASP \cite{kres1}. Results are summarized in Table \ref{tab:info_si}.
For BP and BAs, the AB$^{'}$ stacking is the lowest energy configuration across all D3 methods.
The AB stacking is the next most favorable configuration, with a maximum energy difference of 19 meV.
For BN, the AB$^{'}$ stacking is also most stable with the TS method, while GD3 and BJD3 favor the AB configuration.
The AB, AB$^{'}$, and AA$^{'}$ stacking-orders of the BN compound are notably stable, with minimal energy differences.
Note that these results are for bulk structures and may not apply to two-dimensional structures.
It is worth noting that the interlayer distance correlates directly with the total energy of BX materials. The smaller the interlayer distance, the lower the energy.
\begin{table*}[h!]
\centering
\caption{Total energy variation, $E$, and distance between layers, $d$. The values correspond to the GD3 (BJD3/TS) methods used for dispersion interactions. The dashes indicate unstable geometries.}
\label{tab:info_si}
\resizebox{0.95\textwidth}{!}{
\begin{tabular}{lrccrccrcc}
    \hline
    & \multicolumn{2}{c}{BN}& \multicolumn{2}{c}{BP}& \multicolumn{2}{c}{BAs}\\
    & $E$ [meV] & $d$ [\AA] & $E$ [meV] & $d$ [\AA] & $E$ [meV] & $d$ [\AA] \\
    \hline
 AA    & 60~(62/63)    & 3.52 (3.50/3.44) &  171~(191/158) & 3.98 (3.80/3.97) & ~196~(207/176)~&  4.02~(3.92/4.02) \\
 AA$'$ & 4~(3/13)      & 3.23 (3.19/3.32) &  130~(149/122) & 3.76 (3.60/3.83) & ~144~(163/138)~  &  3.85~(3.79/3.95)  \\
 AB    & 0~(0/5)       & 3.20 (3.17/3.31) &   16~(13/16)   & 3.46 (3.37/3.61) & ~16~(12/19)~  &  3.55~(3.46/3.70) \\
 AB$'$ & 10~(10/0)     & 3.19 (3.21/3.27) &    0~(0/0)     & 3.43 (3.36/3.61) & ~0~(0/0)~    &  3.52~(3.45/3.69) \\
 A$'$B & ---(---/52)   & --- (---/3.45)   &   85~(75/173)  & 3.51 (3.44/3.68) & ~93~(75/81)~  &  3.60~(3.54/3.75) \\\hline
\end{tabular}
}
\end{table*}

\subsection{Vibrational Mode Analysis of Cubic BX Compounds}
\label{sec:cbx}
\begin{figure}[ht!]
    \centering
 \includegraphics[width=0.9\linewidth]{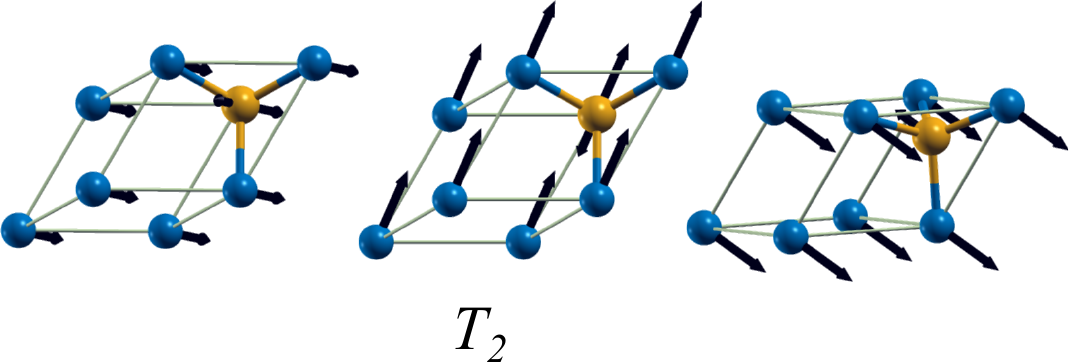}
    \caption{Schematic representations of atomic displacements of $T_2$ Raman-active mode. Orange and blue balls represent X (where X = N, P, and As) and B atoms, respectively. The arrows indicate displacements, derived from the real part of the eigenvectors at the $\Gamma$-point, with their sizes and orientations reflecting the magnitudes and directions of the displacement vectors.}
    \label{fig:cubicRAMAN}
\end{figure}

\begin{table*}[h!]
\centering
\begin{tabular}{|c|c|c|c|c|c|c|c|}
\hline
    \(T_d\) & \(E\) & \(8C_3\) & \(3C_2\) & \(6S_4\) & \(6\sigma_d\) & Linear, Rot. & Quadratic \\ \hline
$A_1$ & 1 & 1 & 1 & 1 & 1 &  & \(x^2+y^2+z^2\) \\ \hline
$A_2$ & 1 & 1 & 1 & -1 & -1 &  &  \\ \hline
$E$ & 2 & -1 & 2 & 0 & 0 &  & \((2z^2-x^2-y^2, x^2-y^2)\) \\ \hline
$T_1$ & 3 & 0 & -1 & 1 & -1 & \((R_x, R_y, R_z)\) &  \\ \hline
$T_2$ & 3 & 0 & -1 & -1 & 1 & \((x, y, z)\) & \((xy, xz, yz)\) \\ \hline
\end{tabular}
\caption{Character table for \(T_d\) point group.}
\label{tab:charc-Td}
\end{table*}
The character table for $T_d$ symmetry is shown in Table \ref{tab:charc-Td}. Each row represents an irreducible representation of the point group (e.g., $A_1$, $A_2$, $E$, $T_1$, $T_2$), while each column corresponds to a symmetry operation or class (e.g., $E$ for the identity, 3$C_3$ for three 120° rotations, 2$C_2$ for two 180° rotations, 4$S_4$ for four 90° improper rotations, and $\sigma_d$ for dihedral mirror planes).
The linear terms ($x$, $y$, $z$) in the character table indicate how basis functions transform under point group symmetry operations, which is crucial for identifying IR-active modes. In $T_d$ symmetry, only $T_2$ contains the linear terms necessary for IR-activity. The quadratic terms ($x^2$, $y^2$, $z^2$, $xy$, $xz$, $yz$) are vital for identifying Raman-active modes. Thus, $A_1$, $E$, and $T_2$ are Raman-active.
Figure \ref{fig:cubicRAMAN} illustrates the atomic displacement and vibration of the $T_2$ Raman-active mode. The $T_2$ mode is a triple-degenerate active mode where the B and X atoms vibrate oppositely along axial directions.

\subsection{Vibrational Mode Analysis of Wurtzite BX Compounds}
\label{sec:wbx}
\begin{table*}[ht!]
\centering
\begin{tabular}{|c|c|c|c|c|c|c|c|c|}
\hline
\(C_{6v}\) & \(E\) & \(2C_6\) & \(2C_3\) & \(C_2\) & \(3\sigma_v\) & \(3\sigma_d\) & Linear, Rot. & Quadratic \\ \hline
$A_1$ & 1 & 1 & 1 & 1 & 1 & 1 &  & \(x^2 + y^2, z^2\) \\ \hline
$A_2$ & 1 & 1 & 1 & 1 & -1 & -1 & \(R_z\) &  \\ \hline
$B_1$ & 1 & -1 & 1 & -1 & 1 & -1 &  &  \\ \hline
$B_2$ & 1 & -1 & 1 & -1 & -1 & 1 &  &  \\ \hline
$E_1$ & 2 & 1 & -1 & -2 & 0 & 0 & \((R_x, R_y)\) & \((xz, yz)\) \\ \hline
$E_2$ & 2 & -1 & -1 & 2 & 0 & 0 & \((x, y)\) & \((x^2 - y^2, xy)\) \\ \hline
\end{tabular}
\caption{Character table for \(C_{6v}\) point group.}
\label{tab:charc-C6v}
\end{table*}

\begin{figure}
    \centering
 \includegraphics[width=0.85\linewidth]{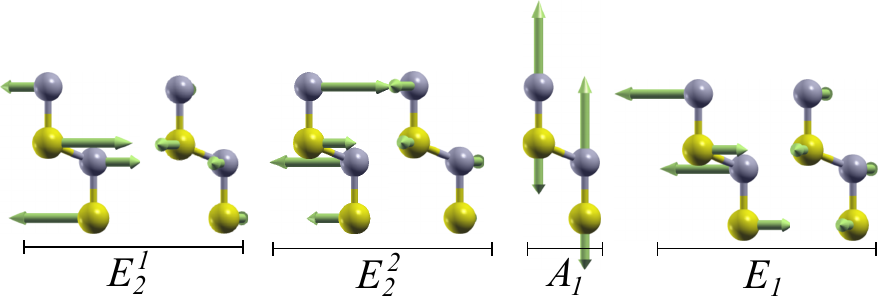}
    \caption{Schematic representations of atomic displacements for Raman-active modes. Yellow and purple balls represent X (where N, P, and As) and B atoms, respectively. The arrows indicate the displacements, derived from the real part of the eigenvectors at the $\Gamma$-point, with their sizes and orientations reflecting the magnitudes and directions of the displacement vectors.}
    \label{fig:wbxmodes}
\end{figure}

The character table for $C_{6v}$ symmetry is shown in Table \ref{tab:charc-C6v}.
Unlike $S_4$, $C_6$ includes two additional symmetry operations: $2C_6$ two 60° rotations and $\sigma_v$ vertical mirror planes.
$E_2$ representation indicates that it is IR-active.
While, $A_1$, $E_1$, and $E_2$ representations are Raman-active modes.
Figure \ref{fig:wbxmodes} illustrates the atomic displacements of the Raman active modes.
In the $E_1$ mode, the B and X atoms vibrate in opposite directions along one of these in-plane dimensions. The $E_2$ mode also involves in-plane vibrations but with distinct symmetry properties; the B and X atoms move oppositely within the basal plane, typically representing non-polar vibrations. In the out-of-plane $A_1$ mode the B and X atoms move in opposite directions along the $c$-axis, perpendicular to the basal plane of the wurtzite structure, and is marked by longitudinal vibrations.

\subsection{Vibrational Mode Analysis of Hexagonal BX Compounds}
\label{sec:hbx}
\begin{table*}[ht!]
\centering
\resizebox{0.97\textwidth}{!}{
\begin{tabular}{|c|c|c|c|c|c|c|c|c|c|c|c|c|c|c|}
\hline 
\(D_{6h}\) & \(E\) & \(2C_6\) & \(2C_3\) & \(C_2\) & \(3C_2'\) & \(3C_2''\) & \(i\) & \(2S_3\) & \(2S_6\) & \(\sigma_h\) & \(3\sigma_d\) & \(3\sigma_v\) & Linear, Rot. & Quadratic \\ \hline
$A_{1g}$ & 1 & 1 & 1 & 1 & 1 & 1 & 1 & 1 & 1 & 1 & 1 & 1 & & \(x^2 + y^2, z^2\) \\ \hline
$A_{2g}$ & 1 & 1 & 1 & 1 & -1 & -1 & 1 & 1 & 1 & 1 & -1 & -1 & \(R_z\) & \\ \hline
$B_{1g}$ & 1 & -1 & 1 & -1 & 1 & -1 & 1 & -1 & 1 & -1 & 1 & -1 & & \\ \hline
$B_{2g}$ & 1 & -1 & 1 & -1 & -1 & 1 & 1 & -1 & 1 & -1 & -1 & 1 & & \\ \hline
$E_{1g}$ & 2 & 1 & -1 & -2 & 0 & 0 & 2 & -1 & -1 & -2 & 0 & 0 & \(R_x, R_y\) & \(xz, yz\) \\ \hline
$E_{2g}$ & 2 & -1 & -1 & 2 & 0 & 0 & 2 & 1 & -1 & 2 & 0 & 0 & & \(x^2 - y^2, xy\) \\ \hline
$A_{1u}$ & 1 & 1 & 1 & 1 & 1 & 1 & -1 & -1 & -1 & -1 & -1 & -1 & & \\ \hline
$A_{2u}$ & 1 & 1 & 1 & 1 & -1 & -1 & -1 & -1 & -1 & -1 & 1 & 1 & \(z\) & \\ \hline
$B_{1u}$ & 1 & -1 & 1 & -1 & 1 & -1 & -1 & 1 & -1 & -1 & -1 & 1 & & \\ \hline
$B_{2u}$ & 1 & -1 & 1 & -1 & -1 & 1 & -1 & 1 & -1 & -1 & 1 & -1 & & \\ \hline
$E_{1u}$ & 2 & 1 & -1 & -2 & 0 & 0 & -2 & -1 & 1 & -2 & 0 & 0 & \(x, y\) & \\ \hline
$E_{2u}$ & 2 & -1 & -1 & 2 & 0 & 0 & -2 & 1 & 1 & -2 & 0 & 0 & & \\ \hline
\end{tabular}
}
\caption{Character table for \(D_{6h}\) point group}
\label{tab:charc-d6h}
\end{table*}
Table \ref{tab:charc-d6h} presents the character table of \(D_{6h}\) point group: $E$ (identity), $2C_6$ (two 60° rotations about the principal axis), $2C_3$ (two 120° rotations), $C_2$ (180° rotation), $3C_2'$ (three 180° rotations about axes perpendicular to the principal axis), $3C_2''$ (three 180° rotations in the horizontal plane), $i$ (inversion through the center of symmetry), $2S_3$ (two 120° improper rotations), $2S_6$ (two 60° improper rotations), $\sigma_h$ (reflection through the horizontal mirror plane), $3\sigma_d$ (reflection through three dihedral mirror planes), and $3\sigma_v$ (reflection through three vertical mirror planes).

\begin{figure}
    \centering
 \includegraphics[width=0.85\linewidth]{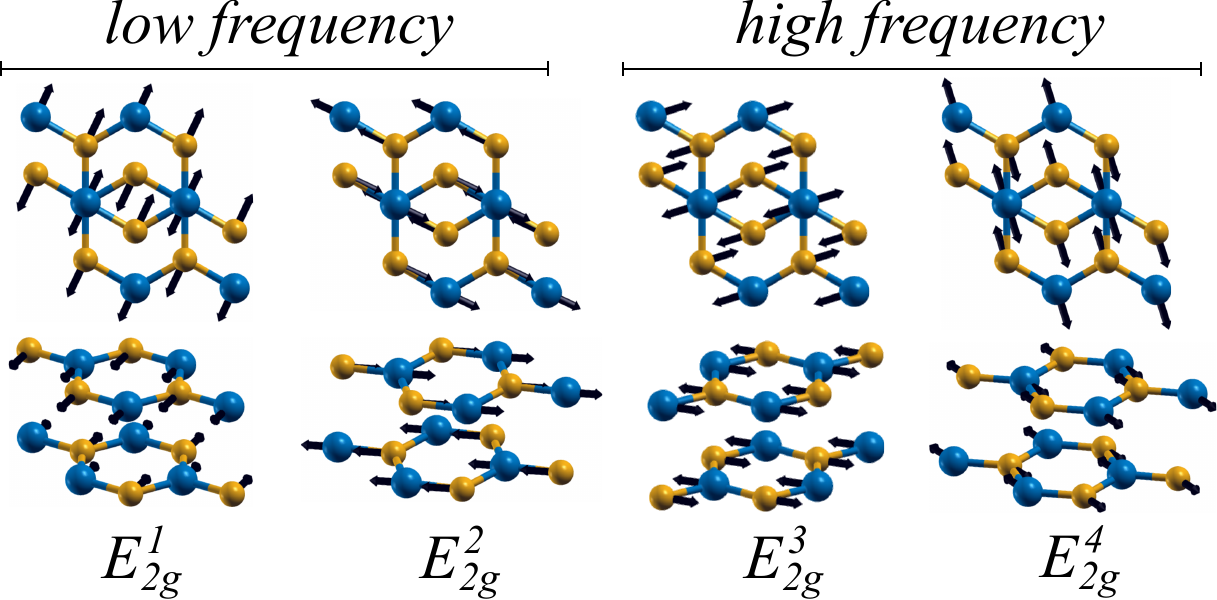}
    \caption{Schematic representations of atomic displacements for Raman-active modes. Orange and blue balls represent X (where X = N, P, As) and B atoms, respectively. The arrows, proportional to the displacements from the real part of the eigenvectors at the $\Gamma$-point, indicate both the magnitude and direction of the displacement vectors.}
    \label{fig:hbx-mode}
\end{figure}

The Raman-active modes are $A_{1g}$, $E_{1g}$, and $E_{2g}$. Only $E_{2g}$-symmetry modes are active in hexagonal BN. Figure \ref{fig:hbx-mode} illustrates the atomic displacements of the low-frequency modes and high-frequency modes.
The low frequency $E_{2g}$ mode is a doubly degenerate vibrational mode denoted as $E_{2g}^1$ and $E_{2g}^2$. In this mode, the B and X atoms within each plane vibrate in the same direction, while adjacent planes vibrate in opposite directions.

The high frequency $E_{2g}$ mode, which is doubly degenerate ($E_{2g}^3$ and $E_{2g}^4$), occurs at a higher frequency. In this mode, the B and X atoms vibrate in opposite directions within the same plane. The bonding forces constrain these in-plane vibrations, leading to higher frequencies.

\subsection{DFT Raman Activity vs. RGDOS Spectrum}

\begin{figure}
    \centering
 \includegraphics[width=0.85\linewidth]{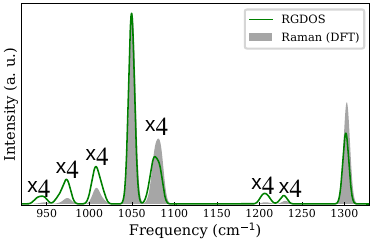}
    \caption{The direct Raman and RGDOS spectra for natural $c-^{\rm nat}$BN compounds. The main peaks are illustrated without magnitude, while the other minor peaks are magnified for a better comparison}
    \label{fig:dftvsrgdos}
\end{figure}

\subsection{Computational Details}
To directly calculate Raman activity, we first computed the phonons with PHONOPY \cite{TOGO2015}, including non-analytical corrections (NAC) to model long-range interactions accurately. Next, we formatted the vibrational modes for VASP to calculate Raman intensities using Eq. (4), which involves derivatives of the macroscopic dielectric tensor with respect to normal mode coordinates.
To impose the isotope effect, we distributed naturally abundant B atoms in a $3\times3\times3$ supercell. 
This approach ensures an accurate capture of vibrational properties and their contributions to the Raman spectrum.

\subsection{Results}
Figure \ref{fig:dftvsrgdos} demonstrates that the peaks resulting from the RGDOS method are well aligned with the direct DFT calculations for $c-$$^{\rm nat}$BN. The main peaks of the Raman spectrum are shown without scaling the magnitude, indicating that the significant vibrational modes are accurately represented. Additional minor peaks were magnified four times larger for a better comparison, highlighting the detailed consistency between the direct calculation and the RGDOS, even for less prominent features. This alignment underscores the reliability and accuracy of the Raman intensity using RGDOS approach.


\subsection{Partial RGDOS of BX Compounds}

\begin{figure*}[ht!]
\centering
\includegraphics[width=0.95\linewidth]{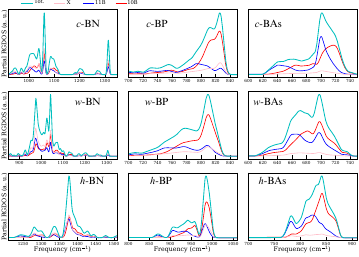}
\caption{The partial RGDOS spectra of $^{11}$B$_{50}$ $^{10}$B$_{50}$X compounds (where X = N, P, or As) in various structures.}
\label{fig:partialRGDOS}
\end{figure*}

The partial RGDOS spectra for $^{11}$B$_{50}$$^{10}$B$_{50}$X compounds are shown in Figure \ref{fig:partialRGDOS}. It details the contributions of different isotopes and atoms to the vibrational modes.

The contributions of $^{10}$B and $^{11}$B are nearly equal for all BN compounds.
In BP, $^{10}$B contributes more to the vibrational modes than $^{11}$B. This indicates that the lighter $^{10}$B isotope is more active in the vibrational modes, resulting in a greater contribution to the RGDOS spectra.
Boron isotopes contribute differently to the vibrational modes of BAs depending on the crystal structure; $^{10}$B and $^{11}$B both equally shape the peaks, however, $^{11}$B is the primary contributor in the lower-frequency and $^{10}$B in higher-frequency shoulder.
Overall, the contribution of the X atom (N, P, or As) varies with the mass and the structure of the compound.

\bibliography{main.bib}

\begin{thebibliography}{67}%
\makeatletter
\providecommand \@ifxundefined [1]{%
 \@ifx{#1\undefined}
}%
\providecommand \@ifnum [1]{%
 \ifnum #1\expandafter \@firstoftwo
 \else \expandafter \@secondoftwo
 \fi
}%
\providecommand \@ifx [1]{%
 \ifx #1\expandafter \@firstoftwo
 \else \expandafter \@secondoftwo
 \fi
}%
\providecommand \natexlab [1]{#1}%
\providecommand \enquote  [1]{``#1''}%
\providecommand \bibnamefont  [1]{#1}%
\providecommand \bibfnamefont [1]{#1}%
\providecommand \citenamefont [1]{#1}%
\providecommand \href@noop [0]{\@secondoftwo}%
\providecommand \href [0]{\begingroup \@sanitize@url \@href}%
\providecommand \@href[1]{\@@startlink{#1}\@@href}%
\providecommand \@@href[1]{\endgroup#1\@@endlink}%
\providecommand \@sanitize@url [0]{\catcode `\\12\catcode `\$12\catcode
  `\&12\catcode `\#12\catcode `\^12\catcode `\_12\catcode `\%12\relax}%
\providecommand \@@startlink[1]{}%
\providecommand \@@endlink[0]{}%
\providecommand \url  [0]{\begingroup\@sanitize@url \@url }%
\providecommand \@url [1]{\endgroup\@href {#1}{\urlprefix }}%
\providecommand \urlprefix  [0]{URL }%
\providecommand \Eprint [0]{\href }%
\providecommand \doibase [0]{http://dx.doi.org/}%
\providecommand \selectlanguage [0]{\@gobble}%
\providecommand \bibinfo  [0]{\@secondoftwo}%
\providecommand \bibfield  [0]{\@secondoftwo}%
\providecommand \translation [1]{[#1]}%
\providecommand \BibitemOpen [0]{}%
\providecommand \bibitemStop [0]{}%
\providecommand \bibitemNoStop [0]{.\EOS\space}%
\providecommand \EOS [0]{\spacefactor3000\relax}%
\providecommand \BibitemShut  [1]{\csname bibitem#1\endcsname}%
\let\auto@bib@innerbib\@empty
\bibitem [{\citenamefont {Kang}\ \emph {et~al.}(2019)\citenamefont {Kang},
  \citenamefont {Li}, \citenamefont {Wu}, \citenamefont {Nguyen},\ and\
  \citenamefont {Hu}}]{Kang2019}%
  \BibitemOpen
  \bibfield  {author} {\bibinfo {author} {\bibfnamefont {J.~S.}\ \bibnamefont
  {Kang}}, \bibinfo {author} {\bibfnamefont {M.}~\bibnamefont {Li}}, \bibinfo
  {author} {\bibfnamefont {H.}~\bibnamefont {Wu}}, \bibinfo {author}
  {\bibfnamefont {H.}~\bibnamefont {Nguyen}}, \ and\ \bibinfo {author}
  {\bibfnamefont {Y.}~\bibnamefont {Hu}},\ }\href {\doibase 10.1063/1.5116025}
  {\bibfield  {journal} {\bibinfo  {journal} {Applied Physics Letters}\
  }\textbf {\bibinfo {volume} {115}},\ \bibinfo {pages} {122103} (\bibinfo
  {year} {2019})}\BibitemShut {NoStop}%
\bibitem [{\citenamefont {Kang}\ \emph {et~al.}(2021)\citenamefont {Kang},
  \citenamefont {Li}, \citenamefont {Wu}, \citenamefont {Nguyen}, \citenamefont
  {Aoki},\ and\ \citenamefont {Hu}}]{Kang2021}%
  \BibitemOpen
  \bibfield  {author} {\bibinfo {author} {\bibfnamefont {J.~S.}\ \bibnamefont
  {Kang}}, \bibinfo {author} {\bibfnamefont {M.}~\bibnamefont {Li}}, \bibinfo
  {author} {\bibfnamefont {H.}~\bibnamefont {Wu}}, \bibinfo {author}
  {\bibfnamefont {H.}~\bibnamefont {Nguyen}}, \bibinfo {author} {\bibfnamefont
  {T.}~\bibnamefont {Aoki}}, \ and\ \bibinfo {author} {\bibfnamefont
  {Y.}~\bibnamefont {Hu}},\ }\href {\doibase 10.1038/s41928-021-00595-9}
  {\bibfield  {journal} {\bibinfo  {journal} {Nature Electronics}\ }\textbf
  {\bibinfo {volume} {4}},\ \bibinfo {pages} {416} (\bibinfo {year}
  {2021})}\BibitemShut {NoStop}%
\bibitem [{\citenamefont {Lee}\ \emph {et~al.}(2024)\citenamefont {Lee},
  \citenamefont {Kim},\ and\ \citenamefont {Kang}}]{Lee_apl_2024}%
  \BibitemOpen
  \bibfield  {author} {\bibinfo {author} {\bibfnamefont {D.}~\bibnamefont
  {Lee}}, \bibinfo {author} {\bibfnamefont {J.}~\bibnamefont {Kim}}, \ and\
  \bibinfo {author} {\bibfnamefont {J.~S.}\ \bibnamefont {Kang}},\ }\href
  {\doibase 10.1063/5.0208339} {\bibfield  {journal} {\bibinfo  {journal}
  {Applied Physics Letters}\ }\textbf {\bibinfo {volume} {124}},\ \bibinfo
  {pages} {222201} (\bibinfo {year} {2024})}\BibitemShut {NoStop}%
\bibitem [{\citenamefont {Mirkarimi}\ \emph {et~al.}(1997)\citenamefont
  {Mirkarimi}, \citenamefont {McCarty},\ and\ \citenamefont
  {Medlin}}]{Mirkarimi1997MSER}%
  \BibitemOpen
  \bibfield  {author} {\bibinfo {author} {\bibfnamefont {P.}~\bibnamefont
  {Mirkarimi}}, \bibinfo {author} {\bibfnamefont {K.}~\bibnamefont {McCarty}},
  \ and\ \bibinfo {author} {\bibfnamefont {D.}~\bibnamefont {Medlin}},\ }\href
  {\doibase https://doi.org/10.1016/S0927-796X(97)00009-0} {\bibfield
  {journal} {\bibinfo  {journal} {Materials Science and Engineering: R:
  Reports}\ }\textbf {\bibinfo {volume} {21}},\ \bibinfo {pages} {47} (\bibinfo
  {year} {1997})}\BibitemShut {NoStop}%
\bibitem [{\citenamefont {Solozhenko}\ and\ \citenamefont
  {Matar}(2022)}]{Solozhenko2022_rsc}%
  \BibitemOpen
  \bibfield  {author} {\bibinfo {author} {\bibfnamefont {V.~L.}\ \bibnamefont
  {Solozhenko}}\ and\ \bibinfo {author} {\bibfnamefont {S.~F.}\ \bibnamefont
  {Matar}},\ }\href {\doibase 10.1039/D2TC00363E} {\bibfield  {journal}
  {\bibinfo  {journal} {J. Mater. Chem. C}\ }\textbf {\bibinfo {volume} {10}},\
  \bibinfo {pages} {3937} (\bibinfo {year} {2022})}\BibitemShut {NoStop}%
\bibitem [{\citenamefont {al-jalali-wal-ikram Shaik}\ and\ \citenamefont
  {Palla}(2021)}]{Shaik2021}%
  \BibitemOpen
  \bibfield  {author} {\bibinfo {author} {\bibfnamefont {A.~B.~D.}\
  \bibnamefont {al-jalali-wal-ikram Shaik}}\ and\ \bibinfo {author}
  {\bibfnamefont {P.}~\bibnamefont {Palla}},\ }\href {\doibase
  10.1038/s41598-021-90804-4} {\bibfield  {journal} {\bibinfo  {journal}
  {Scientific Reports}\ }\textbf {\bibinfo {volume} {11}},\ \bibinfo {pages}
  {12285} (\bibinfo {year} {2021})}\BibitemShut {NoStop}%
\bibitem [{\citenamefont {Ramsay}\ \emph {et~al.}(2023)\citenamefont {Ramsay},
  \citenamefont {Hekmati}, \citenamefont {Patrickson}, \citenamefont {Baber},
  \citenamefont {Arvidsson-Shukur}, \citenamefont {Bennett},\ and\
  \citenamefont {Luxmoore}}]{Ramsay2023}%
  \BibitemOpen
  \bibfield  {author} {\bibinfo {author} {\bibfnamefont {A.~J.}\ \bibnamefont
  {Ramsay}}, \bibinfo {author} {\bibfnamefont {R.}~\bibnamefont {Hekmati}},
  \bibinfo {author} {\bibfnamefont {C.~J.}\ \bibnamefont {Patrickson}},
  \bibinfo {author} {\bibfnamefont {S.}~\bibnamefont {Baber}}, \bibinfo
  {author} {\bibfnamefont {D.~R.~M.}\ \bibnamefont {Arvidsson-Shukur}},
  \bibinfo {author} {\bibfnamefont {A.~J.}\ \bibnamefont {Bennett}}, \ and\
  \bibinfo {author} {\bibfnamefont {I.~J.}\ \bibnamefont {Luxmoore}},\ }\href
  {\doibase 10.1038/s41467-023-36196-7} {\bibfield  {journal} {\bibinfo
  {journal} {Nature Communications}\ }\textbf {\bibinfo {volume} {14}},\
  \bibinfo {pages} {461} (\bibinfo {year} {2023})}\BibitemShut {NoStop}%
\bibitem [{\citenamefont {Pelliciari}\ \emph {et~al.}(2024)\citenamefont
  {Pelliciari}, \citenamefont {Mejia}, \citenamefont {Woods}, \citenamefont
  {Gu}, \citenamefont {Li}, \citenamefont {Chand}, \citenamefont {Fan},
  \citenamefont {Watanabe}, \citenamefont {Taniguchi}, \citenamefont
  {Bisogni},\ and\ \citenamefont {Grosso}}]{Pelliciari2024}%
  \BibitemOpen
  \bibfield  {author} {\bibinfo {author} {\bibfnamefont {J.}~\bibnamefont
  {Pelliciari}}, \bibinfo {author} {\bibfnamefont {E.}~\bibnamefont {Mejia}},
  \bibinfo {author} {\bibfnamefont {J.~M.}\ \bibnamefont {Woods}}, \bibinfo
  {author} {\bibfnamefont {Y.}~\bibnamefont {Gu}}, \bibinfo {author}
  {\bibfnamefont {J.}~\bibnamefont {Li}}, \bibinfo {author} {\bibfnamefont
  {S.~B.}\ \bibnamefont {Chand}}, \bibinfo {author} {\bibfnamefont
  {S.}~\bibnamefont {Fan}}, \bibinfo {author} {\bibfnamefont {K.}~\bibnamefont
  {Watanabe}}, \bibinfo {author} {\bibfnamefont {T.}~\bibnamefont {Taniguchi}},
  \bibinfo {author} {\bibfnamefont {V.}~\bibnamefont {Bisogni}}, \ and\
  \bibinfo {author} {\bibfnamefont {G.}~\bibnamefont {Grosso}},\ }\href
  {\doibase 10.1038/s41563-024-01866-4} {\bibfield  {journal} {\bibinfo
  {journal} {Nature Materials}\ } (\bibinfo {year} {2024}),\
  10.1038/s41563-024-01866-4}\BibitemShut {NoStop}%
\bibitem [{\citenamefont {Kang}\ \emph {et~al.}(2018)\citenamefont {Kang},
  \citenamefont {Li}, \citenamefont {Wu}, \citenamefont {Nguyen},\ and\
  \citenamefont {Hu}}]{Kang2018}%
  \BibitemOpen
  \bibfield  {author} {\bibinfo {author} {\bibfnamefont {J.~S.}\ \bibnamefont
  {Kang}}, \bibinfo {author} {\bibfnamefont {M.}~\bibnamefont {Li}}, \bibinfo
  {author} {\bibfnamefont {H.}~\bibnamefont {Wu}}, \bibinfo {author}
  {\bibfnamefont {H.}~\bibnamefont {Nguyen}}, \ and\ \bibinfo {author}
  {\bibfnamefont {Y.}~\bibnamefont {Hu}},\ }\href {\doibase
  10.1126/science.aat5522} {\bibfield  {journal} {\bibinfo  {journal}
  {Science}\ }\textbf {\bibinfo {volume} {361}},\ \bibinfo {pages} {575}
  (\bibinfo {year} {2018})}\BibitemShut {NoStop}%
\bibitem [{\citenamefont {Li}\ \emph {et~al.}(2018)\citenamefont {Li},
  \citenamefont {Zheng}, \citenamefont {Lv}, \citenamefont {Liu}, \citenamefont
  {Wang}, \citenamefont {Huang}, \citenamefont {Cahill},\ and\ \citenamefont
  {Lv}}]{Li_sci2018}%
  \BibitemOpen
  \bibfield  {author} {\bibinfo {author} {\bibfnamefont {S.}~\bibnamefont
  {Li}}, \bibinfo {author} {\bibfnamefont {Q.}~\bibnamefont {Zheng}}, \bibinfo
  {author} {\bibfnamefont {Y.}~\bibnamefont {Lv}}, \bibinfo {author}
  {\bibfnamefont {X.}~\bibnamefont {Liu}}, \bibinfo {author} {\bibfnamefont
  {X.}~\bibnamefont {Wang}}, \bibinfo {author} {\bibfnamefont {P.~Y.}\
  \bibnamefont {Huang}}, \bibinfo {author} {\bibfnamefont {D.~G.}\ \bibnamefont
  {Cahill}}, \ and\ \bibinfo {author} {\bibfnamefont {B.}~\bibnamefont {Lv}},\
  }\href {\doibase 10.1126/science.aat8982} {\bibfield  {journal} {\bibinfo
  {journal} {Science}\ }\textbf {\bibinfo {volume} {361}},\ \bibinfo {pages}
  {579} (\bibinfo {year} {2018})}\BibitemShut {NoStop}%
\bibitem [{\citenamefont {Mou}\ \emph {et~al.}(2019)\citenamefont {Mou},
  \citenamefont {Wu}, \citenamefont {Xie}, \citenamefont {Zhang}, \citenamefont
  {Ji}, \citenamefont {Huang}, \citenamefont {Wang}, \citenamefont {Luo},
  \citenamefont {Xiong}, \citenamefont {Tang},\ and\ \citenamefont
  {Sun}}]{Mou2019}%
  \BibitemOpen
  \bibfield  {author} {\bibinfo {author} {\bibfnamefont {S.}~\bibnamefont
  {Mou}}, \bibinfo {author} {\bibfnamefont {T.}~\bibnamefont {Wu}}, \bibinfo
  {author} {\bibfnamefont {J.}~\bibnamefont {Xie}}, \bibinfo {author}
  {\bibfnamefont {Y.}~\bibnamefont {Zhang}}, \bibinfo {author} {\bibfnamefont
  {L.}~\bibnamefont {Ji}}, \bibinfo {author} {\bibfnamefont {H.}~\bibnamefont
  {Huang}}, \bibinfo {author} {\bibfnamefont {T.}~\bibnamefont {Wang}},
  \bibinfo {author} {\bibfnamefont {Y.}~\bibnamefont {Luo}}, \bibinfo {author}
  {\bibfnamefont {X.}~\bibnamefont {Xiong}}, \bibinfo {author} {\bibfnamefont
  {B.}~\bibnamefont {Tang}}, \ and\ \bibinfo {author} {\bibfnamefont
  {X.}~\bibnamefont {Sun}},\ }\href {\doibase
  https://doi.org/10.1002/adma.201903499} {\bibfield  {journal} {\bibinfo
  {journal} {Advanced Materials}\ }\textbf {\bibinfo {volume} {31}},\ \bibinfo
  {pages} {1903499} (\bibinfo {year} {2019})}\BibitemShut {NoStop}%
\bibitem [{\citenamefont {Zhang}\ \emph {et~al.}(2023)\citenamefont {Zhang},
  \citenamefont {Zhang}, \citenamefont {Tian}, \citenamefont {Guo},\ and\
  \citenamefont {Chu}}]{Zhang2023_DT}%
  \BibitemOpen
  \bibfield  {author} {\bibinfo {author} {\bibfnamefont {N.}~\bibnamefont
  {Zhang}}, \bibinfo {author} {\bibfnamefont {G.}~\bibnamefont {Zhang}},
  \bibinfo {author} {\bibfnamefont {Y.}~\bibnamefont {Tian}}, \bibinfo {author}
  {\bibfnamefont {Y.}~\bibnamefont {Guo}}, \ and\ \bibinfo {author}
  {\bibfnamefont {K.}~\bibnamefont {Chu}},\ }\href {\doibase
  10.1039/D3DT00551H} {\bibfield  {journal} {\bibinfo  {journal} {Dalton
  Trans.}\ }\textbf {\bibinfo {volume} {52}},\ \bibinfo {pages} {4290}
  (\bibinfo {year} {2023})}\BibitemShut {NoStop}%
\bibitem [{\citenamefont {Vuong}\ \emph {et~al.}(2018)\citenamefont {Vuong},
  \citenamefont {Liu}, \citenamefont {Van~der Lee}, \citenamefont {Cuscó},
  \citenamefont {Artús}, \citenamefont {Michel}, \citenamefont {Valvin},
  \citenamefont {Edgar}, \citenamefont {Cassabois},\ and\ \citenamefont
  {Gil}}]{Vuong2018}%
  \BibitemOpen
  \bibfield  {author} {\bibinfo {author} {\bibfnamefont {T.~Q.~P.}\
  \bibnamefont {Vuong}}, \bibinfo {author} {\bibfnamefont {S.}~\bibnamefont
  {Liu}}, \bibinfo {author} {\bibfnamefont {A.}~\bibnamefont {Van~der Lee}},
  \bibinfo {author} {\bibfnamefont {R.}~\bibnamefont {Cuscó}}, \bibinfo
  {author} {\bibfnamefont {L.}~\bibnamefont {Artús}}, \bibinfo {author}
  {\bibfnamefont {T.}~\bibnamefont {Michel}}, \bibinfo {author} {\bibfnamefont
  {P.}~\bibnamefont {Valvin}}, \bibinfo {author} {\bibfnamefont {J.~H.}\
  \bibnamefont {Edgar}}, \bibinfo {author} {\bibfnamefont {G.}~\bibnamefont
  {Cassabois}}, \ and\ \bibinfo {author} {\bibfnamefont {B.}~\bibnamefont
  {Gil}},\ }\href {\doibase 10.1038/nmat5048} {\bibfield  {journal} {\bibinfo
  {journal} {Nature Materials}\ }\textbf {\bibinfo {volume} {17}},\ \bibinfo
  {pages} {152} (\bibinfo {year} {2018})}\BibitemShut {NoStop}%
\bibitem [{\citenamefont {Janzen}\ \emph {et~al.}(2024)\citenamefont {Janzen},
  \citenamefont {Schutte}, \citenamefont {Plo}, \citenamefont {Rousseau},
  \citenamefont {Michel}, \citenamefont {Desrat}, \citenamefont {Valvin},
  \citenamefont {Jacques}, \citenamefont {Cassabois}, \citenamefont {Gil},\
  and\ \citenamefont {Edgar}}]{Janzen2024AdvMat}%
  \BibitemOpen
  \bibfield  {author} {\bibinfo {author} {\bibfnamefont {E.}~\bibnamefont
  {Janzen}}, \bibinfo {author} {\bibfnamefont {H.}~\bibnamefont {Schutte}},
  \bibinfo {author} {\bibfnamefont {J.}~\bibnamefont {Plo}}, \bibinfo {author}
  {\bibfnamefont {A.}~\bibnamefont {Rousseau}}, \bibinfo {author}
  {\bibfnamefont {T.}~\bibnamefont {Michel}}, \bibinfo {author} {\bibfnamefont
  {W.}~\bibnamefont {Desrat}}, \bibinfo {author} {\bibfnamefont
  {P.}~\bibnamefont {Valvin}}, \bibinfo {author} {\bibfnamefont
  {V.}~\bibnamefont {Jacques}}, \bibinfo {author} {\bibfnamefont
  {G.}~\bibnamefont {Cassabois}}, \bibinfo {author} {\bibfnamefont
  {B.}~\bibnamefont {Gil}}, \ and\ \bibinfo {author} {\bibfnamefont {J.~H.}\
  \bibnamefont {Edgar}},\ }\href {\doibase
  https://doi.org/10.1002/adma.202306033} {\bibfield  {journal} {\bibinfo
  {journal} {Advanced Materials}\ }\textbf {\bibinfo {volume} {36}},\ \bibinfo
  {pages} {2306033} (\bibinfo {year} {2024})}\BibitemShut {NoStop}%
\bibitem [{\citenamefont {Cuscó}\ \emph {et~al.}(2019)\citenamefont {Cuscó},
  \citenamefont {Edgar}, \citenamefont {Liu}, \citenamefont {Cassabois},
  \citenamefont {Gil},\ and\ \citenamefont {Artús}}]{Cusco_2019}%
  \BibitemOpen
  \bibfield  {author} {\bibinfo {author} {\bibfnamefont {R.}~\bibnamefont
  {Cuscó}}, \bibinfo {author} {\bibfnamefont {J.}~\bibnamefont {Edgar}},
  \bibinfo {author} {\bibfnamefont {S.}~\bibnamefont {Liu}}, \bibinfo {author}
  {\bibfnamefont {G.}~\bibnamefont {Cassabois}}, \bibinfo {author}
  {\bibfnamefont {B.}~\bibnamefont {Gil}}, \ and\ \bibinfo {author}
  {\bibfnamefont {L.}~\bibnamefont {Artús}},\ }\href {\doibase
  10.1088/1361-6463/ab1cab} {\bibfield  {journal} {\bibinfo  {journal} {Journal
  of Physics D: Applied Physics}\ }\textbf {\bibinfo {volume} {52}},\ \bibinfo
  {pages} {303001} (\bibinfo {year} {2019})}\BibitemShut {NoStop}%
\bibitem [{\citenamefont {Chen}\ \emph {et~al.}(2020)\citenamefont {Chen},
  \citenamefont {Song}, \citenamefont {Ravichandran}, \citenamefont {Zheng},
  \citenamefont {Chen}, \citenamefont {Lee}, \citenamefont {Sun}, \citenamefont
  {Li}, \citenamefont {Udalamatta~Gamage}, \citenamefont {Tian}, \citenamefont
  {Ding}, \citenamefont {Song}, \citenamefont {Rai}, \citenamefont {Wu},
  \citenamefont {Koirala}, \citenamefont {Schmidt}, \citenamefont {Watanabe},
  \citenamefont {Lv}, \citenamefont {Ren}, \citenamefont {Shi}, \citenamefont
  {Cahill}, \citenamefont {Taniguchi}, \citenamefont {Broido},\ and\
  \citenamefont {Chen}}]{Chen555}%
  \BibitemOpen
  \bibfield  {author} {\bibinfo {author} {\bibfnamefont {K.}~\bibnamefont
  {Chen}}, \bibinfo {author} {\bibfnamefont {B.}~\bibnamefont {Song}}, \bibinfo
  {author} {\bibfnamefont {N.~K.}\ \bibnamefont {Ravichandran}}, \bibinfo
  {author} {\bibfnamefont {Q.}~\bibnamefont {Zheng}}, \bibinfo {author}
  {\bibfnamefont {X.}~\bibnamefont {Chen}}, \bibinfo {author} {\bibfnamefont
  {H.}~\bibnamefont {Lee}}, \bibinfo {author} {\bibfnamefont {H.}~\bibnamefont
  {Sun}}, \bibinfo {author} {\bibfnamefont {S.}~\bibnamefont {Li}}, \bibinfo
  {author} {\bibfnamefont {G.~A.~G.}\ \bibnamefont {Udalamatta~Gamage}},
  \bibinfo {author} {\bibfnamefont {F.}~\bibnamefont {Tian}}, \bibinfo {author}
  {\bibfnamefont {Z.}~\bibnamefont {Ding}}, \bibinfo {author} {\bibfnamefont
  {Q.}~\bibnamefont {Song}}, \bibinfo {author} {\bibfnamefont {A.}~\bibnamefont
  {Rai}}, \bibinfo {author} {\bibfnamefont {H.}~\bibnamefont {Wu}}, \bibinfo
  {author} {\bibfnamefont {P.}~\bibnamefont {Koirala}}, \bibinfo {author}
  {\bibfnamefont {A.~J.}\ \bibnamefont {Schmidt}}, \bibinfo {author}
  {\bibfnamefont {K.}~\bibnamefont {Watanabe}}, \bibinfo {author}
  {\bibfnamefont {B.}~\bibnamefont {Lv}}, \bibinfo {author} {\bibfnamefont
  {Z.}~\bibnamefont {Ren}}, \bibinfo {author} {\bibfnamefont {L.}~\bibnamefont
  {Shi}}, \bibinfo {author} {\bibfnamefont {D.~G.}\ \bibnamefont {Cahill}},
  \bibinfo {author} {\bibfnamefont {T.}~\bibnamefont {Taniguchi}}, \bibinfo
  {author} {\bibfnamefont {D.}~\bibnamefont {Broido}}, \ and\ \bibinfo {author}
  {\bibfnamefont {G.}~\bibnamefont {Chen}},\ }\href {\doibase
  10.1126/science.aaz6149} {\bibfield  {journal} {\bibinfo  {journal}
  {Science}\ }\textbf {\bibinfo {volume} {367}},\ \bibinfo {pages} {555}
  (\bibinfo {year} {2020})}\BibitemShut {NoStop}%
\bibitem [{\citenamefont {Lindsay}\ and\ \citenamefont
  {Broido}(2011)}]{Lindsay2011}%
  \BibitemOpen
  \bibfield  {author} {\bibinfo {author} {\bibfnamefont {L.}~\bibnamefont
  {Lindsay}}\ and\ \bibinfo {author} {\bibfnamefont {D.~A.}\ \bibnamefont
  {Broido}},\ }\href {\doibase 10.1103/PhysRevB.84.155421} {\bibfield
  {journal} {\bibinfo  {journal} {Phys. Rev. B}\ }\textbf {\bibinfo {volume}
  {84}},\ \bibinfo {pages} {155421} (\bibinfo {year} {2011})}\BibitemShut
  {NoStop}%
\bibitem [{\citenamefont {Vuong}(2018)}]{vuong_thesis}%
  \BibitemOpen
  \bibfield  {author} {\bibinfo {author} {\bibfnamefont {P.}~\bibnamefont
  {Vuong}},\ }\emph {\bibinfo {title} {{Optical spectroscopy of boron nitride
  heterostructures}}},\ \href {https://tel.archives-ouvertes.fr/tel-02152566}
  {\bibinfo {type} {Theses}},\ \bibinfo  {school} {{Universit{\'e}
  Montpellier}} (\bibinfo {year} {2018})\BibitemShut {NoStop}%
\bibitem [{\citenamefont {Yoshikawa}(2023)}]{Yoshikawa2023}%
  \BibitemOpen
  \bibfield  {author} {\bibinfo {author} {\bibfnamefont {M.}~\bibnamefont
  {Yoshikawa}},\ }\enquote {\bibinfo {title} {Raman and infrared (ir)
  spectroscopy},}\ in\ \href {\doibase 10.1007/978-3-031-19722-2_2} {\emph
  {\bibinfo {booktitle} {Advanced Optical Spectroscopy Techniques for
  Semiconductors: Raman, Infrared, and Cathodoluminescence Spectroscopy}}}\
  (\bibinfo  {publisher} {Springer International Publishing},\ \bibinfo
  {address} {Cham},\ \bibinfo {year} {2023})\ pp.\ \bibinfo {pages}
  {3--25}\BibitemShut {NoStop}%
\bibitem [{\citenamefont {Cardona}\ and\ \citenamefont
  {G{\"u}ntherodt}(1975)}]{cardona1975}%
  \BibitemOpen
  \bibfield  {author} {\bibinfo {author} {\bibfnamefont {M.}~\bibnamefont
  {Cardona}}\ and\ \bibinfo {author} {\bibfnamefont {G.}~\bibnamefont
  {G{\"u}ntherodt}},\ }\href {https://books.google.fi/books?id=QPtAAQAAIAAJ}
  {\emph {\bibinfo {title} {Light Scattering in Solids}}},\ \bibinfo {series}
  {Light Scattering in Solids}\ No.\ \bibinfo {number} {v. 1-7; v. 9}\
  (\bibinfo  {publisher} {Springer-Verlag},\ \bibinfo {year}
  {1975})\BibitemShut {NoStop}%
\bibitem [{\citenamefont {Werninghaus}\ \emph {et~al.}(1997)\citenamefont
  {Werninghaus}, \citenamefont {Friedrich}, \citenamefont {Hahn}, \citenamefont
  {Richter},\ and\ \citenamefont {Zahn}}]{Werninghaus1997}%
  \BibitemOpen
  \bibfield  {author} {\bibinfo {author} {\bibfnamefont {T.}~\bibnamefont
  {Werninghaus}}, \bibinfo {author} {\bibfnamefont {M.}~\bibnamefont
  {Friedrich}}, \bibinfo {author} {\bibfnamefont {J.}~\bibnamefont {Hahn}},
  \bibinfo {author} {\bibfnamefont {F.}~\bibnamefont {Richter}}, \ and\
  \bibinfo {author} {\bibfnamefont {D.}~\bibnamefont {Zahn}},\ }\href {\doibase
  https://doi.org/10.1016/S0925-9635(96)00645-0} {\bibfield  {journal}
  {\bibinfo  {journal} {Diamond and Related Materials}\ }\textbf {\bibinfo
  {volume} {6}},\ \bibinfo {pages} {612} (\bibinfo {year} {1997})}\BibitemShut
  {NoStop}%
\bibitem [{\citenamefont {Sachdev}\ \emph {et~al.}(1997)\citenamefont
  {Sachdev}, \citenamefont {Haubner}, \citenamefont {Nöth},\ and\
  \citenamefont {Lux}}]{Sachdev1997}%
  \BibitemOpen
  \bibfield  {author} {\bibinfo {author} {\bibfnamefont {H.}~\bibnamefont
  {Sachdev}}, \bibinfo {author} {\bibfnamefont {R.}~\bibnamefont {Haubner}},
  \bibinfo {author} {\bibfnamefont {H.}~\bibnamefont {Nöth}}, \ and\ \bibinfo
  {author} {\bibfnamefont {B.}~\bibnamefont {Lux}},\ }\href {\doibase
  https://doi.org/10.1016/S0925-9635(96)00697-8} {\bibfield  {journal}
  {\bibinfo  {journal} {Diamond and Related Materials}\ }\textbf {\bibinfo
  {volume} {6}},\ \bibinfo {pages} {286} (\bibinfo {year} {1997})}\BibitemShut
  {NoStop}%
\bibitem [{\citenamefont {Reich}\ \emph {et~al.}(2005)\citenamefont {Reich},
  \citenamefont {Ferrari}, \citenamefont {Arenal}, \citenamefont {Loiseau},
  \citenamefont {Bello},\ and\ \citenamefont {Robertson}}]{Reich2005_PRB}%
  \BibitemOpen
  \bibfield  {author} {\bibinfo {author} {\bibfnamefont {S.}~\bibnamefont
  {Reich}}, \bibinfo {author} {\bibfnamefont {A.~C.}\ \bibnamefont {Ferrari}},
  \bibinfo {author} {\bibfnamefont {R.}~\bibnamefont {Arenal}}, \bibinfo
  {author} {\bibfnamefont {A.}~\bibnamefont {Loiseau}}, \bibinfo {author}
  {\bibfnamefont {I.}~\bibnamefont {Bello}}, \ and\ \bibinfo {author}
  {\bibfnamefont {J.}~\bibnamefont {Robertson}},\ }\href {\doibase
  10.1103/PhysRevB.71.205201} {\bibfield  {journal} {\bibinfo  {journal} {Phys.
  Rev. B}\ }\textbf {\bibinfo {volume} {71}},\ \bibinfo {pages} {205201}
  (\bibinfo {year} {2005})}\BibitemShut {NoStop}%
\bibitem [{\citenamefont {Narayan}\ and\ \citenamefont
  {Bhaumik}(2016)}]{Jagdish2016}%
  \BibitemOpen
  \bibfield  {author} {\bibinfo {author} {\bibfnamefont {J.}~\bibnamefont
  {Narayan}}\ and\ \bibinfo {author} {\bibfnamefont {A.}~\bibnamefont
  {Bhaumik}},\ }\href {\doibase 10.1063/1.4941095} {\bibfield  {journal}
  {\bibinfo  {journal} {APL Materials}\ }\textbf {\bibinfo {volume} {4}},\
  \bibinfo {pages} {020701} (\bibinfo {year} {2016})}\BibitemShut {NoStop}%
\bibitem [{\citenamefont {Griffin}\ \emph {et~al.}(2018)\citenamefont
  {Griffin}, \citenamefont {Harvey}, \citenamefont {Cunningham}, \citenamefont
  {Scullion}, \citenamefont {Tian}, \citenamefont {Shih}, \citenamefont
  {Gruening}, \citenamefont {Donegan}, \citenamefont {Santos}, \citenamefont
  {Backes},\ and\ \citenamefont {Coleman}}]{Griffin2018}%
  \BibitemOpen
  \bibfield  {author} {\bibinfo {author} {\bibfnamefont {A.}~\bibnamefont
  {Griffin}}, \bibinfo {author} {\bibfnamefont {A.}~\bibnamefont {Harvey}},
  \bibinfo {author} {\bibfnamefont {B.}~\bibnamefont {Cunningham}}, \bibinfo
  {author} {\bibfnamefont {D.}~\bibnamefont {Scullion}}, \bibinfo {author}
  {\bibfnamefont {T.}~\bibnamefont {Tian}}, \bibinfo {author} {\bibfnamefont
  {C.-J.}\ \bibnamefont {Shih}}, \bibinfo {author} {\bibfnamefont
  {M.}~\bibnamefont {Gruening}}, \bibinfo {author} {\bibfnamefont {J.~F.}\
  \bibnamefont {Donegan}}, \bibinfo {author} {\bibfnamefont {E.~J.~G.}\
  \bibnamefont {Santos}}, \bibinfo {author} {\bibfnamefont {C.}~\bibnamefont
  {Backes}}, \ and\ \bibinfo {author} {\bibfnamefont {J.~N.}\ \bibnamefont
  {Coleman}},\ }\href {\doibase 10.1021/acs.chemmater.7b05188} {\bibfield
  {journal} {\bibinfo  {journal} {Chemistry of Materials}\ }\textbf {\bibinfo
  {volume} {30}},\ \bibinfo {pages} {1998} (\bibinfo {year}
  {2018})}\BibitemShut {NoStop}%
\bibitem [{\citenamefont {Cai}\ \emph {et~al.}(2017)\citenamefont {Cai},
  \citenamefont {Scullion}, \citenamefont {Falin}, \citenamefont {Watanabe},
  \citenamefont {Taniguchi}, \citenamefont {Chen}, \citenamefont {Santos},\
  and\ \citenamefont {Li}}]{Cai2017}%
  \BibitemOpen
  \bibfield  {author} {\bibinfo {author} {\bibfnamefont {Q.}~\bibnamefont
  {Cai}}, \bibinfo {author} {\bibfnamefont {D.}~\bibnamefont {Scullion}},
  \bibinfo {author} {\bibfnamefont {A.}~\bibnamefont {Falin}}, \bibinfo
  {author} {\bibfnamefont {K.}~\bibnamefont {Watanabe}}, \bibinfo {author}
  {\bibfnamefont {T.}~\bibnamefont {Taniguchi}}, \bibinfo {author}
  {\bibfnamefont {Y.}~\bibnamefont {Chen}}, \bibinfo {author} {\bibfnamefont
  {E.~J.~G.}\ \bibnamefont {Santos}}, \ and\ \bibinfo {author} {\bibfnamefont
  {L.~H.}\ \bibnamefont {Li}},\ }\href {\doibase 10.1039/C6NR09312D} {\bibfield
   {journal} {\bibinfo  {journal} {Nanoscale}\ }\textbf {\bibinfo {volume}
  {9}},\ \bibinfo {pages} {3059} (\bibinfo {year} {2017})}\BibitemShut
  {NoStop}%
\bibitem [{\citenamefont {Sachdev}(2003)}]{Sachdev2003_D}%
  \BibitemOpen
  \bibfield  {author} {\bibinfo {author} {\bibfnamefont {H.}~\bibnamefont
  {Sachdev}},\ }\href {\doibase https://doi.org/10.1016/S0925-9635(03)00072-4}
  {\bibfield  {journal} {\bibinfo  {journal} {Diamond and Related Materials}\
  }\textbf {\bibinfo {volume} {12}},\ \bibinfo {pages} {1275} (\bibinfo {year}
  {2003})}\BibitemShut {NoStop}%
\bibitem [{\citenamefont {Solozhenko}\ \emph {et~al.}(2014)\citenamefont
  {Solozhenko}, \citenamefont {Kurakevych}, \citenamefont {Le~Godec},
  \citenamefont {Kurnosov},\ and\ \citenamefont {Oganov}}]{solozhenko2014_JAP}%
  \BibitemOpen
  \bibfield  {author} {\bibinfo {author} {\bibfnamefont {V.~L.}\ \bibnamefont
  {Solozhenko}}, \bibinfo {author} {\bibfnamefont {O.~O.}\ \bibnamefont
  {Kurakevych}}, \bibinfo {author} {\bibfnamefont {Y.}~\bibnamefont
  {Le~Godec}}, \bibinfo {author} {\bibfnamefont {A.~V.}\ \bibnamefont
  {Kurnosov}}, \ and\ \bibinfo {author} {\bibfnamefont {A.~R.}\ \bibnamefont
  {Oganov}},\ }\href {\doibase 10.1063/1.4890231} {\bibfield  {journal}
  {\bibinfo  {journal} {Journal of Applied Physics}\ }\textbf {\bibinfo
  {volume} {116}},\ \bibinfo {pages} {033501} (\bibinfo {year}
  {2014})}\BibitemShut {NoStop}%
\bibitem [{\citenamefont {Berger}\ and\ \citenamefont
  {Komsa}(2024)}]{Berger_prb2024}%
  \BibitemOpen
  \bibfield  {author} {\bibinfo {author} {\bibfnamefont {E.}~\bibnamefont
  {Berger}}\ and\ \bibinfo {author} {\bibfnamefont {H.-P.}\ \bibnamefont
  {Komsa}},\ }\href {\doibase 10.1103/PhysRevMaterials.8.043802} {\bibfield
  {journal} {\bibinfo  {journal} {Phys. Rev. Mater.}\ }\textbf {\bibinfo
  {volume} {8}},\ \bibinfo {pages} {043802} (\bibinfo {year}
  {2024})}\BibitemShut {NoStop}%
\bibitem [{\citenamefont {Hadjiev}\ \emph {et~al.}(2014)\citenamefont
  {Hadjiev}, \citenamefont {Iliev}, \citenamefont {Lv}, \citenamefont {Ren},\
  and\ \citenamefont {Chu}}]{Hadjiev2014_PRB}%
  \BibitemOpen
  \bibfield  {author} {\bibinfo {author} {\bibfnamefont {V.~G.}\ \bibnamefont
  {Hadjiev}}, \bibinfo {author} {\bibfnamefont {M.~N.}\ \bibnamefont {Iliev}},
  \bibinfo {author} {\bibfnamefont {B.}~\bibnamefont {Lv}}, \bibinfo {author}
  {\bibfnamefont {Z.~F.}\ \bibnamefont {Ren}}, \ and\ \bibinfo {author}
  {\bibfnamefont {C.~W.}\ \bibnamefont {Chu}},\ }\href {\doibase
  10.1103/PhysRevB.89.024308} {\bibfield  {journal} {\bibinfo  {journal} {Phys.
  Rev. B}\ }\textbf {\bibinfo {volume} {89}},\ \bibinfo {pages} {024308}
  (\bibinfo {year} {2014})}\BibitemShut {NoStop}%
\bibitem [{\citenamefont {Cataldo}\ and\ \citenamefont
  {Iglesias-Groth}(2017)}]{Cataldo2017}%
  \BibitemOpen
  \bibfield  {author} {\bibinfo {author} {\bibfnamefont {F.}~\bibnamefont
  {Cataldo}}\ and\ \bibinfo {author} {\bibfnamefont {S.}~\bibnamefont
  {Iglesias-Groth}},\ }\href {\doibase 10.1007/s10967-017-5289-8} {\bibfield
  {journal} {\bibinfo  {journal} {Journal of Radioanalytical and Nuclear
  Chemistry}\ }\textbf {\bibinfo {volume} {313}},\ \bibinfo {pages} {261}
  (\bibinfo {year} {2017})}\BibitemShut {NoStop}%
\bibitem [{\citenamefont {Gnatyuk}\ \emph {et~al.}(2024)\citenamefont
  {Gnatyuk}, \citenamefont {Dovbeshko}, \citenamefont {Dementjev},
  \citenamefont {Chernyakova}, \citenamefont {Posudievsky}, \citenamefont
  {Kupchak}, \citenamefont {Kolesnyk}, \citenamefont {Solyanik},\ and\
  \citenamefont {Karpicz}}]{Gnatyuk2024opmat}%
  \BibitemOpen
  \bibfield  {author} {\bibinfo {author} {\bibfnamefont {O.}~\bibnamefont
  {Gnatyuk}}, \bibinfo {author} {\bibfnamefont {G.}~\bibnamefont {Dovbeshko}},
  \bibinfo {author} {\bibfnamefont {A.}~\bibnamefont {Dementjev}}, \bibinfo
  {author} {\bibfnamefont {K.}~\bibnamefont {Chernyakova}}, \bibinfo {author}
  {\bibfnamefont {O.}~\bibnamefont {Posudievsky}}, \bibinfo {author}
  {\bibfnamefont {I.}~\bibnamefont {Kupchak}}, \bibinfo {author} {\bibfnamefont
  {D.}~\bibnamefont {Kolesnyk}}, \bibinfo {author} {\bibfnamefont
  {G.}~\bibnamefont {Solyanik}}, \ and\ \bibinfo {author} {\bibfnamefont
  {R.}~\bibnamefont {Karpicz}},\ }\href {\doibase
  https://doi.org/10.1016/j.omx.2024.100323} {\bibfield  {journal} {\bibinfo
  {journal} {Optical Materials: X}\ }\textbf {\bibinfo {volume} {22}},\
  \bibinfo {pages} {100323} (\bibinfo {year} {2024})}\BibitemShut {NoStop}%
\bibitem [{\citenamefont {He}\ \emph {et~al.}(2021)\citenamefont {He},
  \citenamefont {Lindsay}, \citenamefont {Beechem}, \citenamefont {Folland},
  \citenamefont {Matson}, \citenamefont {Watanabe}, \citenamefont {Zavalin},
  \citenamefont {Ueda}, \citenamefont {Collins}, \citenamefont {Taniguchi}
  \emph {et~al.}}]{he2021phonon}%
  \BibitemOpen
  \bibfield  {author} {\bibinfo {author} {\bibfnamefont {M.}~\bibnamefont
  {He}}, \bibinfo {author} {\bibfnamefont {L.}~\bibnamefont {Lindsay}},
  \bibinfo {author} {\bibfnamefont {T.~E.}\ \bibnamefont {Beechem}}, \bibinfo
  {author} {\bibfnamefont {T.}~\bibnamefont {Folland}}, \bibinfo {author}
  {\bibfnamefont {J.}~\bibnamefont {Matson}}, \bibinfo {author} {\bibfnamefont
  {K.}~\bibnamefont {Watanabe}}, \bibinfo {author} {\bibfnamefont
  {A.}~\bibnamefont {Zavalin}}, \bibinfo {author} {\bibfnamefont
  {A.}~\bibnamefont {Ueda}}, \bibinfo {author} {\bibfnamefont {W.~E.}\
  \bibnamefont {Collins}}, \bibinfo {author} {\bibfnamefont {T.}~\bibnamefont
  {Taniguchi}},  \emph {et~al.},\ }\href@noop {} {\bibfield  {journal}
  {\bibinfo  {journal} {Journal of materials research}\ }\textbf {\bibinfo
  {volume} {36}},\ \bibinfo {pages} {4394} (\bibinfo {year}
  {2021})}\BibitemShut {NoStop}%
\bibitem [{\citenamefont {Zhu}\ \emph {et~al.}(2023)\citenamefont {Zhu},
  \citenamefont {Lu},\ and\ \citenamefont {Zheng}}]{Zhu2023_pss}%
  \BibitemOpen
  \bibfield  {author} {\bibinfo {author} {\bibfnamefont {S.}~\bibnamefont
  {Zhu}}, \bibinfo {author} {\bibfnamefont {X.}~\bibnamefont {Lu}}, \ and\
  \bibinfo {author} {\bibfnamefont {W.}~\bibnamefont {Zheng}},\ }\href
  {\doibase https://doi.org/10.1002/pssb.202200585} {\bibfield  {journal}
  {\bibinfo  {journal} {physica status solidi (b)}\ }\textbf {\bibinfo {volume}
  {260}},\ \bibinfo {pages} {2200585} (\bibinfo {year} {2023})}\BibitemShut
  {NoStop}%
\bibitem [{\citenamefont {Rai}\ \emph {et~al.}(2021)\citenamefont {Rai},
  \citenamefont {Li}, \citenamefont {Wu}, \citenamefont {Lv},\ and\
  \citenamefont {Cahill}}]{Rai_PRM2020}%
  \BibitemOpen
  \bibfield  {author} {\bibinfo {author} {\bibfnamefont {A.}~\bibnamefont
  {Rai}}, \bibinfo {author} {\bibfnamefont {S.}~\bibnamefont {Li}}, \bibinfo
  {author} {\bibfnamefont {H.}~\bibnamefont {Wu}}, \bibinfo {author}
  {\bibfnamefont {B.}~\bibnamefont {Lv}}, \ and\ \bibinfo {author}
  {\bibfnamefont {D.~G.}\ \bibnamefont {Cahill}},\ }\href {\doibase
  10.1103/PhysRevMaterials.5.013603} {\bibfield  {journal} {\bibinfo  {journal}
  {Phys. Rev. Materials}\ }\textbf {\bibinfo {volume} {5}},\ \bibinfo {pages}
  {013603} (\bibinfo {year} {2021})}\BibitemShut {NoStop}%
\bibitem [{\citenamefont {Cusc\'o}\ \emph {et~al.}(2021)\citenamefont
  {Cusc\'o}, \citenamefont {Pellicer-Porres}, \citenamefont {Edgar},
  \citenamefont {Li}, \citenamefont {Segura},\ and\ \citenamefont
  {Art\'us}}]{Cusco2021}%
  \BibitemOpen
  \bibfield  {author} {\bibinfo {author} {\bibfnamefont {R.}~\bibnamefont
  {Cusc\'o}}, \bibinfo {author} {\bibfnamefont {J.}~\bibnamefont
  {Pellicer-Porres}}, \bibinfo {author} {\bibfnamefont {J.~H.}\ \bibnamefont
  {Edgar}}, \bibinfo {author} {\bibfnamefont {J.}~\bibnamefont {Li}}, \bibinfo
  {author} {\bibfnamefont {A.}~\bibnamefont {Segura}}, \ and\ \bibinfo {author}
  {\bibfnamefont {L.}~\bibnamefont {Art\'us}},\ }\href {\doibase
  10.1103/PhysRevB.103.085204} {\bibfield  {journal} {\bibinfo  {journal}
  {Phys. Rev. B}\ }\textbf {\bibinfo {volume} {103}},\ \bibinfo {pages}
  {085204} (\bibinfo {year} {2021})}\BibitemShut {NoStop}%
\bibitem [{\citenamefont {Hashemi}\ \emph {et~al.}(2019)\citenamefont
  {Hashemi}, \citenamefont {Krasheninnikov}, \citenamefont {Puska},\ and\
  \citenamefont {Komsa}}]{Hashemi2019_PRM}%
  \BibitemOpen
  \bibfield  {author} {\bibinfo {author} {\bibfnamefont {A.}~\bibnamefont
  {Hashemi}}, \bibinfo {author} {\bibfnamefont {A.~V.}\ \bibnamefont
  {Krasheninnikov}}, \bibinfo {author} {\bibfnamefont {M.}~\bibnamefont
  {Puska}}, \ and\ \bibinfo {author} {\bibfnamefont {H.-P.}\ \bibnamefont
  {Komsa}},\ }\href {\doibase 10.1103/PhysRevMaterials.3.023806} {\bibfield
  {journal} {\bibinfo  {journal} {Phys. Rev. Materials}\ }\textbf {\bibinfo
  {volume} {3}},\ \bibinfo {pages} {023806} (\bibinfo {year}
  {2019})}\BibitemShut {NoStop}%
\bibitem [{\citenamefont {Kou}\ \emph {et~al.}(2020)\citenamefont {Kou},
  \citenamefont {Hashemi}, \citenamefont {Puska}, \citenamefont
  {Krasheninnikov},\ and\ \citenamefont {Komsa}}]{Kou2020}%
  \BibitemOpen
  \bibfield  {author} {\bibinfo {author} {\bibfnamefont {Z.}~\bibnamefont
  {Kou}}, \bibinfo {author} {\bibfnamefont {A.}~\bibnamefont {Hashemi}},
  \bibinfo {author} {\bibfnamefont {M.~J.}\ \bibnamefont {Puska}}, \bibinfo
  {author} {\bibfnamefont {A.~V.}\ \bibnamefont {Krasheninnikov}}, \ and\
  \bibinfo {author} {\bibfnamefont {H.-P.}\ \bibnamefont {Komsa}},\ }\href
  {\doibase 10.1038/s41524-020-0320-y} {\bibfield  {journal} {\bibinfo
  {journal} {npj Computational Materials}\ }\textbf {\bibinfo {volume} {6}},\
  \bibinfo {pages} {59} (\bibinfo {year} {2020})}\BibitemShut {NoStop}%
\bibitem [{\citenamefont {Kresse}\ and\ \citenamefont {Hafner}(1993)}]{kres1}%
  \BibitemOpen
  \bibfield  {author} {\bibinfo {author} {\bibfnamefont {G.}~\bibnamefont
  {Kresse}}\ and\ \bibinfo {author} {\bibfnamefont {J.}~\bibnamefont
  {Hafner}},\ }\href {\doibase 10.1103/PhysRevB.47.558} {\bibfield  {journal}
  {\bibinfo  {journal} {Phys. Rev. B}\ }\textbf {\bibinfo {volume} {47}},\
  \bibinfo {pages} {558} (\bibinfo {year} {1993})}\BibitemShut {NoStop}%
\bibitem [{\citenamefont {Bl\"ochl}(1994)}]{PAW1994}%
  \BibitemOpen
  \bibfield  {author} {\bibinfo {author} {\bibfnamefont {P.~E.}\ \bibnamefont
  {Bl\"ochl}},\ }\href {\doibase 10.1103/PhysRevB.50.17953} {\bibfield
  {journal} {\bibinfo  {journal} {Phys. Rev. B}\ }\textbf {\bibinfo {volume}
  {50}},\ \bibinfo {pages} {17953} (\bibinfo {year} {1994})}\BibitemShut
  {NoStop}%
\bibitem [{\citenamefont {Perdew}\ \emph {et~al.}(2008)\citenamefont {Perdew},
  \citenamefont {Ruzsinszky}, \citenamefont {Csonka}, \citenamefont {Vydrov},
  \citenamefont {Scuseria}, \citenamefont {Constantin}, \citenamefont {Zhou},\
  and\ \citenamefont {Burke}}]{PBEsol2008}%
  \BibitemOpen
  \bibfield  {author} {\bibinfo {author} {\bibfnamefont {J.}~\bibnamefont
  {Perdew}}, \bibinfo {author} {\bibfnamefont {A.}~\bibnamefont {Ruzsinszky}},
  \bibinfo {author} {\bibfnamefont {G.}~\bibnamefont {Csonka}}, \bibinfo
  {author} {\bibfnamefont {O.}~\bibnamefont {Vydrov}}, \bibinfo {author}
  {\bibfnamefont {G.}~\bibnamefont {Scuseria}}, \bibinfo {author}
  {\bibfnamefont {L.}~\bibnamefont {Constantin}}, \bibinfo {author}
  {\bibfnamefont {X.}~\bibnamefont {Zhou}}, \ and\ \bibinfo {author}
  {\bibfnamefont {K.}~\bibnamefont {Burke}},\ }\href {\doibase
  10.1103/PhysRevLett.100.136406} {\bibfield  {journal} {\bibinfo  {journal}
  {Physical Review Letters}\ }\textbf {\bibinfo {volume} {100}} (\bibinfo
  {year} {2008}),\ 10.1103/PhysRevLett.100.136406}\BibitemShut {NoStop}%
\bibitem [{\citenamefont {Perdew}\ \emph {et~al.}(1996)\citenamefont {Perdew},
  \citenamefont {Burke},\ and\ \citenamefont {Ernzerhof}}]{PBE1996PRL}%
  \BibitemOpen
  \bibfield  {author} {\bibinfo {author} {\bibfnamefont {J.~P.}\ \bibnamefont
  {Perdew}}, \bibinfo {author} {\bibfnamefont {K.}~\bibnamefont {Burke}}, \
  and\ \bibinfo {author} {\bibfnamefont {M.}~\bibnamefont {Ernzerhof}},\ }\href
  {\doibase 10.1103/PhysRevLett.77.3865} {\bibfield  {journal} {\bibinfo
  {journal} {Phys. Rev. Lett.}\ }\textbf {\bibinfo {volume} {77}},\ \bibinfo
  {pages} {3865} (\bibinfo {year} {1996})}\BibitemShut {NoStop}%
\bibitem [{\citenamefont {Grimme}\ \emph {et~al.}(2010)\citenamefont {Grimme},
  \citenamefont {Antony}, \citenamefont {Ehrlich},\ and\ \citenamefont
  {Krieg}}]{Grimme2010_D3}%
  \BibitemOpen
  \bibfield  {author} {\bibinfo {author} {\bibfnamefont {S.}~\bibnamefont
  {Grimme}}, \bibinfo {author} {\bibfnamefont {J.}~\bibnamefont {Antony}},
  \bibinfo {author} {\bibfnamefont {S.}~\bibnamefont {Ehrlich}}, \ and\
  \bibinfo {author} {\bibfnamefont {H.}~\bibnamefont {Krieg}},\ }\href
  {\doibase 10.1063/1.3382344} {\bibfield  {journal} {\bibinfo  {journal} {The
  Journal of Chemical Physics}\ }\textbf {\bibinfo {volume} {132}},\ \bibinfo
  {pages} {154104} (\bibinfo {year} {2010})}\BibitemShut {NoStop}%
\bibitem [{\citenamefont {Grimme}\ \emph {et~al.}(2011)\citenamefont {Grimme},
  \citenamefont {Ehrlich},\ and\ \citenamefont {Goerigk}}]{Grimme2011_JCC}%
  \BibitemOpen
  \bibfield  {author} {\bibinfo {author} {\bibfnamefont {S.}~\bibnamefont
  {Grimme}}, \bibinfo {author} {\bibfnamefont {S.}~\bibnamefont {Ehrlich}}, \
  and\ \bibinfo {author} {\bibfnamefont {L.}~\bibnamefont {Goerigk}},\ }\href
  {\doibase https://doi.org/10.1002/jcc.21759} {\bibfield  {journal} {\bibinfo
  {journal} {Journal of Computational Chemistry}\ }\textbf {\bibinfo {volume}
  {32}},\ \bibinfo {pages} {1456} (\bibinfo {year} {2011})}\BibitemShut
  {NoStop}%
\bibitem [{\citenamefont {Tkatchenko}\ and\ \citenamefont
  {Scheffler}(2009)}]{Tkatchenkoprb_2009}%
  \BibitemOpen
  \bibfield  {author} {\bibinfo {author} {\bibfnamefont {A.}~\bibnamefont
  {Tkatchenko}}\ and\ \bibinfo {author} {\bibfnamefont {M.}~\bibnamefont
  {Scheffler}},\ }\href {\doibase 10.1103/PhysRevLett.102.073005} {\bibfield
  {journal} {\bibinfo  {journal} {Phys. Rev. Lett.}\ }\textbf {\bibinfo
  {volume} {102}},\ \bibinfo {pages} {073005} (\bibinfo {year}
  {2009})}\BibitemShut {NoStop}%
\bibitem [{\citenamefont {Togo}\ and\ \citenamefont {Tanaka}(2015)}]{TOGO2015}%
  \BibitemOpen
  \bibfield  {author} {\bibinfo {author} {\bibfnamefont {A.}~\bibnamefont
  {Togo}}\ and\ \bibinfo {author} {\bibfnamefont {I.}~\bibnamefont {Tanaka}},\
  }\href {\doibase https://doi.org/10.1016/j.scriptamat.2015.07.021} {\bibfield
   {journal} {\bibinfo  {journal} {Scripta Materialia}\ }\textbf {\bibinfo
  {volume} {108}},\ \bibinfo {pages} {1 } (\bibinfo {year} {2015})}\BibitemShut
  {NoStop}%
\bibitem [{\citenamefont {Pick}\ \emph {et~al.}(1970)\citenamefont {Pick},
  \citenamefont {Cohen},\ and\ \citenamefont {Martin}}]{Pick1970_PRB}%
  \BibitemOpen
  \bibfield  {author} {\bibinfo {author} {\bibfnamefont {R.~M.}\ \bibnamefont
  {Pick}}, \bibinfo {author} {\bibfnamefont {M.~H.}\ \bibnamefont {Cohen}}, \
  and\ \bibinfo {author} {\bibfnamefont {R.~M.}\ \bibnamefont {Martin}},\
  }\href {\doibase 10.1103/PhysRevB.1.910} {\bibfield  {journal} {\bibinfo
  {journal} {Phys. Rev. B}\ }\textbf {\bibinfo {volume} {1}},\ \bibinfo {pages}
  {910} (\bibinfo {year} {1970})}\BibitemShut {NoStop}%
\bibitem [{\citenamefont {Popper}\ and\ \citenamefont
  {Ingles}(1957)}]{Popper1957nat}%
  \BibitemOpen
  \bibfield  {author} {\bibinfo {author} {\bibfnamefont {P.}~\bibnamefont
  {Popper}}\ and\ \bibinfo {author} {\bibfnamefont {T.~A.}\ \bibnamefont
  {Ingles}},\ }\href {\doibase 10.1038/1791075a0} {\bibfield  {journal}
  {\bibinfo  {journal} {Nature}\ }\textbf {\bibinfo {volume} {179}},\ \bibinfo
  {pages} {1075} (\bibinfo {year} {1957})}\BibitemShut {NoStop}%
\bibitem [{\citenamefont {Kang}\ \emph {et~al.}(2017)\citenamefont {Kang},
  \citenamefont {Wu},\ and\ \citenamefont {Hu}}]{Kang2017_nl}%
  \BibitemOpen
  \bibfield  {author} {\bibinfo {author} {\bibfnamefont {J.~S.}\ \bibnamefont
  {Kang}}, \bibinfo {author} {\bibfnamefont {H.}~\bibnamefont {Wu}}, \ and\
  \bibinfo {author} {\bibfnamefont {Y.}~\bibnamefont {Hu}},\ }\href {\doibase
  10.1021/acs.nanolett.7b03437} {\bibfield  {journal} {\bibinfo  {journal}
  {Nano Letters}\ }\textbf {\bibinfo {volume} {17}},\ \bibinfo {pages} {7507}
  (\bibinfo {year} {2017})},\ \bibinfo {note} {pMID: 29115845}\BibitemShut
  {NoStop}%
\bibitem [{\citenamefont {Warner}\ \emph {et~al.}(2010)\citenamefont {Warner},
  \citenamefont {Rümmeli}, \citenamefont {Bachmatiuk},\ and\ \citenamefont
  {Büchner}}]{Warner2010_acs_nano}%
  \BibitemOpen
  \bibfield  {author} {\bibinfo {author} {\bibfnamefont {J.~H.}\ \bibnamefont
  {Warner}}, \bibinfo {author} {\bibfnamefont {M.~H.}\ \bibnamefont
  {Rümmeli}}, \bibinfo {author} {\bibfnamefont {A.}~\bibnamefont
  {Bachmatiuk}}, \ and\ \bibinfo {author} {\bibfnamefont {B.}~\bibnamefont
  {Büchner}},\ }\href {\doibase 10.1021/nn901648q} {\bibfield  {journal}
  {\bibinfo  {journal} {ACS Nano}\ }\textbf {\bibinfo {volume} {4}},\ \bibinfo
  {pages} {1299} (\bibinfo {year} {2010})}\BibitemShut {NoStop}%
\bibitem [{\citenamefont {Yasuda}\ \emph {et~al.}(2021)\citenamefont {Yasuda},
  \citenamefont {Wang}, \citenamefont {Watanabe}, \citenamefont {Taniguchi},\
  and\ \citenamefont {Jarillo-Herrero}}]{Kenji2021sci}%
  \BibitemOpen
  \bibfield  {author} {\bibinfo {author} {\bibfnamefont {K.}~\bibnamefont
  {Yasuda}}, \bibinfo {author} {\bibfnamefont {X.}~\bibnamefont {Wang}},
  \bibinfo {author} {\bibfnamefont {K.}~\bibnamefont {Watanabe}}, \bibinfo
  {author} {\bibfnamefont {T.}~\bibnamefont {Taniguchi}}, \ and\ \bibinfo
  {author} {\bibfnamefont {P.}~\bibnamefont {Jarillo-Herrero}},\ }\href
  {\doibase 10.1126/science.abd3230} {\bibfield  {journal} {\bibinfo  {journal}
  {Science}\ }\textbf {\bibinfo {volume} {372}},\ \bibinfo {pages} {1458}
  (\bibinfo {year} {2021})}\BibitemShut {NoStop}%
\bibitem [{\citenamefont {Luo}\ \emph {et~al.}(2015)\citenamefont {Luo},
  \citenamefont {Lu}, \citenamefont {Cong}, \citenamefont {Yu}, \citenamefont
  {Xiong},\ and\ \citenamefont {Quek}}]{Luo2015sci}%
  \BibitemOpen
  \bibfield  {author} {\bibinfo {author} {\bibfnamefont {X.}~\bibnamefont
  {Luo}}, \bibinfo {author} {\bibfnamefont {X.}~\bibnamefont {Lu}}, \bibinfo
  {author} {\bibfnamefont {C.}~\bibnamefont {Cong}}, \bibinfo {author}
  {\bibfnamefont {T.}~\bibnamefont {Yu}}, \bibinfo {author} {\bibfnamefont
  {Q.}~\bibnamefont {Xiong}}, \ and\ \bibinfo {author} {\bibfnamefont {S.~Y.}\
  \bibnamefont {Quek}},\ }\href {\doibase 10.1038/srep14565} {\bibfield
  {journal} {\bibinfo  {journal} {Scientific Reports}\ }\textbf {\bibinfo
  {volume} {5}},\ \bibinfo {pages} {14565} (\bibinfo {year}
  {2015})}\BibitemShut {NoStop}%
\bibitem [{\citenamefont {Gilbert}\ \emph {et~al.}(2019)\citenamefont
  {Gilbert}, \citenamefont {Pham}, \citenamefont {Dogan}, \citenamefont {Oh},
  \citenamefont {Shevitski}, \citenamefont {Schumm}, \citenamefont {Liu},
  \citenamefont {Ercius}, \citenamefont {Aloni}, \citenamefont {Cohen},\ and\
  \citenamefont {Zettl}}]{Gilbert_2019}%
  \BibitemOpen
  \bibfield  {author} {\bibinfo {author} {\bibfnamefont {S.~M.}\ \bibnamefont
  {Gilbert}}, \bibinfo {author} {\bibfnamefont {T.}~\bibnamefont {Pham}},
  \bibinfo {author} {\bibfnamefont {M.}~\bibnamefont {Dogan}}, \bibinfo
  {author} {\bibfnamefont {S.}~\bibnamefont {Oh}}, \bibinfo {author}
  {\bibfnamefont {B.}~\bibnamefont {Shevitski}}, \bibinfo {author}
  {\bibfnamefont {G.}~\bibnamefont {Schumm}}, \bibinfo {author} {\bibfnamefont
  {S.}~\bibnamefont {Liu}}, \bibinfo {author} {\bibfnamefont {P.}~\bibnamefont
  {Ercius}}, \bibinfo {author} {\bibfnamefont {S.}~\bibnamefont {Aloni}},
  \bibinfo {author} {\bibfnamefont {M.~L.}\ \bibnamefont {Cohen}}, \ and\
  \bibinfo {author} {\bibfnamefont {A.}~\bibnamefont {Zettl}},\ }\href
  {\doibase 10.1088/2053-1583/ab0e24} {\bibfield  {journal} {\bibinfo
  {journal} {2D Materials}\ }\textbf {\bibinfo {volume} {6}},\ \bibinfo {pages}
  {021006} (\bibinfo {year} {2019})}\BibitemShut {NoStop}%
\bibitem [{\citenamefont {Constantinescu}\ \emph {et~al.}(2013)\citenamefont
  {Constantinescu}, \citenamefont {Kuc},\ and\ \citenamefont
  {Heine}}]{Constantinescu2013-PRL}%
  \BibitemOpen
  \bibfield  {author} {\bibinfo {author} {\bibfnamefont {G.}~\bibnamefont
  {Constantinescu}}, \bibinfo {author} {\bibfnamefont {A.}~\bibnamefont {Kuc}},
  \ and\ \bibinfo {author} {\bibfnamefont {T.}~\bibnamefont {Heine}},\ }\href
  {\doibase 10.1103/PhysRevLett.111.036104} {\bibfield  {journal} {\bibinfo
  {journal} {Phys. Rev. Lett.}\ }\textbf {\bibinfo {volume} {111}},\ \bibinfo
  {pages} {036104} (\bibinfo {year} {2013})}\BibitemShut {NoStop}%
\bibitem [{\citenamefont {Marom}\ \emph {et~al.}(2010)\citenamefont {Marom},
  \citenamefont {Bernstein}, \citenamefont {Garel}, \citenamefont {Tkatchenko},
  \citenamefont {Joselevich}, \citenamefont {Kronik},\ and\ \citenamefont
  {Hod}}]{MaromPRL2010}%
  \BibitemOpen
  \bibfield  {author} {\bibinfo {author} {\bibfnamefont {N.}~\bibnamefont
  {Marom}}, \bibinfo {author} {\bibfnamefont {J.}~\bibnamefont {Bernstein}},
  \bibinfo {author} {\bibfnamefont {J.}~\bibnamefont {Garel}}, \bibinfo
  {author} {\bibfnamefont {A.}~\bibnamefont {Tkatchenko}}, \bibinfo {author}
  {\bibfnamefont {E.}~\bibnamefont {Joselevich}}, \bibinfo {author}
  {\bibfnamefont {L.}~\bibnamefont {Kronik}}, \ and\ \bibinfo {author}
  {\bibfnamefont {O.}~\bibnamefont {Hod}},\ }\href {\doibase
  10.1103/PhysRevLett.105.046801} {\bibfield  {journal} {\bibinfo  {journal}
  {Phys. Rev. Lett.}\ }\textbf {\bibinfo {volume} {105}},\ \bibinfo {pages}
  {046801} (\bibinfo {year} {2010})}\BibitemShut {NoStop}%
\bibitem [{\citenamefont {Henck}\ \emph {et~al.}(2017)\citenamefont {Henck},
  \citenamefont {Pierucci}, \citenamefont {Fugallo}, \citenamefont {Avila},
  \citenamefont {Cassabois}, \citenamefont {Dappe}, \citenamefont {Silly},
  \citenamefont {Chen}, \citenamefont {Gil}, \citenamefont {Gatti},
  \citenamefont {Sottile}, \citenamefont {Sirotti}, \citenamefont {Asensio},\
  and\ \citenamefont {Ouerghi}}]{Henck2017_PRB}%
  \BibitemOpen
  \bibfield  {author} {\bibinfo {author} {\bibfnamefont {H.}~\bibnamefont
  {Henck}}, \bibinfo {author} {\bibfnamefont {D.}~\bibnamefont {Pierucci}},
  \bibinfo {author} {\bibfnamefont {G.}~\bibnamefont {Fugallo}}, \bibinfo
  {author} {\bibfnamefont {J.}~\bibnamefont {Avila}}, \bibinfo {author}
  {\bibfnamefont {G.}~\bibnamefont {Cassabois}}, \bibinfo {author}
  {\bibfnamefont {Y.~J.}\ \bibnamefont {Dappe}}, \bibinfo {author}
  {\bibfnamefont {M.~G.}\ \bibnamefont {Silly}}, \bibinfo {author}
  {\bibfnamefont {C.}~\bibnamefont {Chen}}, \bibinfo {author} {\bibfnamefont
  {B.}~\bibnamefont {Gil}}, \bibinfo {author} {\bibfnamefont {M.}~\bibnamefont
  {Gatti}}, \bibinfo {author} {\bibfnamefont {F.}~\bibnamefont {Sottile}},
  \bibinfo {author} {\bibfnamefont {F.}~\bibnamefont {Sirotti}}, \bibinfo
  {author} {\bibfnamefont {M.~C.}\ \bibnamefont {Asensio}}, \ and\ \bibinfo
  {author} {\bibfnamefont {A.}~\bibnamefont {Ouerghi}},\ }\href {\doibase
  10.1103/PhysRevB.95.085410} {\bibfield  {journal} {\bibinfo  {journal} {Phys.
  Rev. B}\ }\textbf {\bibinfo {volume} {95}},\ \bibinfo {pages} {085410}
  (\bibinfo {year} {2017})}\BibitemShut {NoStop}%
\bibitem [{\citenamefont {Sanjurjo}\ \emph {et~al.}(1983)\citenamefont
  {Sanjurjo}, \citenamefont {L\'opez-Cruz}, \citenamefont {Vogl},\ and\
  \citenamefont {Cardona}}]{Sanjurjo83_PRB}%
  \BibitemOpen
  \bibfield  {author} {\bibinfo {author} {\bibfnamefont {J.~A.}\ \bibnamefont
  {Sanjurjo}}, \bibinfo {author} {\bibfnamefont {E.}~\bibnamefont
  {L\'opez-Cruz}}, \bibinfo {author} {\bibfnamefont {P.}~\bibnamefont {Vogl}},
  \ and\ \bibinfo {author} {\bibfnamefont {M.}~\bibnamefont {Cardona}},\ }\href
  {\doibase 10.1103/PhysRevB.28.4579} {\bibfield  {journal} {\bibinfo
  {journal} {Phys. Rev. B}\ }\textbf {\bibinfo {volume} {28}},\ \bibinfo
  {pages} {4579} (\bibinfo {year} {1983})}\BibitemShut {NoStop}%
\bibitem [{\citenamefont {Serrano}\ \emph {et~al.}(2007)\citenamefont
  {Serrano}, \citenamefont {Bosak}, \citenamefont {Arenal}, \citenamefont
  {Krisch}, \citenamefont {Watanabe}, \citenamefont {Taniguchi}, \citenamefont
  {Kanda}, \citenamefont {Rubio},\ and\ \citenamefont {Wirtz}}]{Serrano2007}%
  \BibitemOpen
  \bibfield  {author} {\bibinfo {author} {\bibfnamefont {J.}~\bibnamefont
  {Serrano}}, \bibinfo {author} {\bibfnamefont {A.}~\bibnamefont {Bosak}},
  \bibinfo {author} {\bibfnamefont {R.}~\bibnamefont {Arenal}}, \bibinfo
  {author} {\bibfnamefont {M.}~\bibnamefont {Krisch}}, \bibinfo {author}
  {\bibfnamefont {K.}~\bibnamefont {Watanabe}}, \bibinfo {author}
  {\bibfnamefont {T.}~\bibnamefont {Taniguchi}}, \bibinfo {author}
  {\bibfnamefont {H.}~\bibnamefont {Kanda}}, \bibinfo {author} {\bibfnamefont
  {A.}~\bibnamefont {Rubio}}, \ and\ \bibinfo {author} {\bibfnamefont
  {L.}~\bibnamefont {Wirtz}},\ }\href {\doibase 10.1103/PhysRevLett.98.095503}
  {\bibfield  {journal} {\bibinfo  {journal} {Phys. Rev. Lett.}\ }\textbf
  {\bibinfo {volume} {98}},\ \bibinfo {pages} {095503} (\bibinfo {year}
  {2007})}\BibitemShut {NoStop}%
\bibitem [{\citenamefont {Mignuzzi}\ \emph {et~al.}(2015)\citenamefont
  {Mignuzzi}, \citenamefont {Pollard}, \citenamefont {Bonini}, \citenamefont
  {Brennan}, \citenamefont {Gilmore}, \citenamefont {Pimenta}, \citenamefont
  {Richards},\ and\ \citenamefont {Roy}}]{Mignuzzi2015}%
  \BibitemOpen
  \bibfield  {author} {\bibinfo {author} {\bibfnamefont {S.}~\bibnamefont
  {Mignuzzi}}, \bibinfo {author} {\bibfnamefont {A.~J.}\ \bibnamefont
  {Pollard}}, \bibinfo {author} {\bibfnamefont {N.}~\bibnamefont {Bonini}},
  \bibinfo {author} {\bibfnamefont {B.}~\bibnamefont {Brennan}}, \bibinfo
  {author} {\bibfnamefont {I.~S.}\ \bibnamefont {Gilmore}}, \bibinfo {author}
  {\bibfnamefont {M.~A.}\ \bibnamefont {Pimenta}}, \bibinfo {author}
  {\bibfnamefont {D.}~\bibnamefont {Richards}}, \ and\ \bibinfo {author}
  {\bibfnamefont {D.}~\bibnamefont {Roy}},\ }\href {\doibase
  10.1103/PhysRevB.91.195411} {\bibfield  {journal} {\bibinfo  {journal} {Phys.
  Rev. B}\ }\textbf {\bibinfo {volume} {91}},\ \bibinfo {pages} {195411}
  (\bibinfo {year} {2015})}\BibitemShut {NoStop}%
\bibitem [{\citenamefont {Richter}\ \emph {et~al.}(1981)\citenamefont
  {Richter}, \citenamefont {Wang},\ and\ \citenamefont {Ley}}]{Richter1981}%
  \BibitemOpen
  \bibfield  {author} {\bibinfo {author} {\bibfnamefont {H.}~\bibnamefont
  {Richter}}, \bibinfo {author} {\bibfnamefont {Z.}~\bibnamefont {Wang}}, \
  and\ \bibinfo {author} {\bibfnamefont {L.}~\bibnamefont {Ley}},\ }\href
  {\doibase https://doi.org/10.1016/0038-1098(81)90337-9} {\bibfield  {journal}
  {\bibinfo  {journal} {Solid State Communications}\ }\textbf {\bibinfo
  {volume} {39}},\ \bibinfo {pages} {625} (\bibinfo {year} {1981})}\BibitemShut
  {NoStop}%
\bibitem [{\citenamefont {Haque}\ and\ \citenamefont
  {Narayan}(2021)}]{Haque2021_ACS}%
  \BibitemOpen
  \bibfield  {author} {\bibinfo {author} {\bibfnamefont {A.}~\bibnamefont
  {Haque}}\ and\ \bibinfo {author} {\bibfnamefont {J.}~\bibnamefont
  {Narayan}},\ }\href {\doibase 10.1021/acsaelm.0c01130} {\bibfield  {journal}
  {\bibinfo  {journal} {ACS Applied Electronic Materials}\ }\textbf {\bibinfo
  {volume} {3}},\ \bibinfo {pages} {1359} (\bibinfo {year} {2021})}\BibitemShut
  {NoStop}%
\bibitem [{\citenamefont {Padavala}\ \emph {et~al.}(2018)\citenamefont
  {Padavala}, \citenamefont {{Al Atabi}}, \citenamefont {Tengdelius},
  \citenamefont {Lu}, \citenamefont {Högberg},\ and\ \citenamefont
  {Edgar}}]{Padavala2018}%
  \BibitemOpen
  \bibfield  {author} {\bibinfo {author} {\bibfnamefont {B.}~\bibnamefont
  {Padavala}}, \bibinfo {author} {\bibfnamefont {H.}~\bibnamefont {{Al
  Atabi}}}, \bibinfo {author} {\bibfnamefont {L.}~\bibnamefont {Tengdelius}},
  \bibinfo {author} {\bibfnamefont {J.}~\bibnamefont {Lu}}, \bibinfo {author}
  {\bibfnamefont {H.}~\bibnamefont {Högberg}}, \ and\ \bibinfo {author}
  {\bibfnamefont {J.}~\bibnamefont {Edgar}},\ }\href {\doibase
  https://doi.org/10.1016/j.jcrysgro.2017.11.014} {\bibfield  {journal}
  {\bibinfo  {journal} {Journal of Crystal Growth}\ }\textbf {\bibinfo {volume}
  {483}},\ \bibinfo {pages} {115} (\bibinfo {year} {2018})}\BibitemShut
  {NoStop}%
\bibitem [{\citenamefont {Zheng}\ \emph {et~al.}(2018)\citenamefont {Zheng},
  \citenamefont {Li}, \citenamefont {Li}, \citenamefont {Lv}, \citenamefont
  {Liu}, \citenamefont {Huang}, \citenamefont {Broido}, \citenamefont {Lv},\
  and\ \citenamefont {Cahill}}]{Zheng2018}%
  \BibitemOpen
  \bibfield  {author} {\bibinfo {author} {\bibfnamefont {Q.}~\bibnamefont
  {Zheng}}, \bibinfo {author} {\bibfnamefont {S.}~\bibnamefont {Li}}, \bibinfo
  {author} {\bibfnamefont {C.}~\bibnamefont {Li}}, \bibinfo {author}
  {\bibfnamefont {Y.}~\bibnamefont {Lv}}, \bibinfo {author} {\bibfnamefont
  {X.}~\bibnamefont {Liu}}, \bibinfo {author} {\bibfnamefont {P.~Y.}\
  \bibnamefont {Huang}}, \bibinfo {author} {\bibfnamefont {D.~A.}\ \bibnamefont
  {Broido}}, \bibinfo {author} {\bibfnamefont {B.}~\bibnamefont {Lv}}, \ and\
  \bibinfo {author} {\bibfnamefont {D.~G.}\ \bibnamefont {Cahill}},\ }\href
  {\doibase https://doi.org/10.1002/adfm.201805116} {\bibfield  {journal}
  {\bibinfo  {journal} {Advanced Functional Materials}\ }\textbf {\bibinfo
  {volume} {28}},\ \bibinfo {pages} {1805116} (\bibinfo {year}
  {2018})}\BibitemShut {NoStop}%
\bibitem [{\citenamefont {Sun}\ \emph {et~al.}(2019)\citenamefont {Sun},
  \citenamefont {Chen}, \citenamefont {Gamage}, \citenamefont {Ziyaee},
  \citenamefont {Wang}, \citenamefont {Wang}, \citenamefont {Hadjiev},
  \citenamefont {Tian}, \citenamefont {Chen},\ and\ \citenamefont
  {Ren}}]{Sun-material2019}%
  \BibitemOpen
  \bibfield  {author} {\bibinfo {author} {\bibfnamefont {H.}~\bibnamefont
  {Sun}}, \bibinfo {author} {\bibfnamefont {K.}~\bibnamefont {Chen}}, \bibinfo
  {author} {\bibfnamefont {G.}~\bibnamefont {Gamage}}, \bibinfo {author}
  {\bibfnamefont {H.}~\bibnamefont {Ziyaee}}, \bibinfo {author} {\bibfnamefont
  {F.}~\bibnamefont {Wang}}, \bibinfo {author} {\bibfnamefont {Y.}~\bibnamefont
  {Wang}}, \bibinfo {author} {\bibfnamefont {V.}~\bibnamefont {Hadjiev}},
  \bibinfo {author} {\bibfnamefont {F.}~\bibnamefont {Tian}}, \bibinfo {author}
  {\bibfnamefont {G.}~\bibnamefont {Chen}}, \ and\ \bibinfo {author}
  {\bibfnamefont {Z.}~\bibnamefont {Ren}},\ }\href {\doibase
  https://doi.org/10.1016/j.mtphys.2019.100169} {\bibfield  {journal} {\bibinfo
   {journal} {Materials Today Physics}\ }\textbf {\bibinfo {volume} {11}},\
  \bibinfo {pages} {100169} (\bibinfo {year} {2019})}\BibitemShut {NoStop}%
\bibitem [{\citenamefont {Yuan}\ \emph {et~al.}(2019)\citenamefont {Yuan},
  \citenamefont {Li}, \citenamefont {Lindsay}, \citenamefont {Cherns},
  \citenamefont {Pomeroy}, \citenamefont {Liu}, \citenamefont {Edgar},\ and\
  \citenamefont {Kuball}}]{Yuan2019}%
  \BibitemOpen
  \bibfield  {author} {\bibinfo {author} {\bibfnamefont {C.}~\bibnamefont
  {Yuan}}, \bibinfo {author} {\bibfnamefont {J.}~\bibnamefont {Li}}, \bibinfo
  {author} {\bibfnamefont {L.}~\bibnamefont {Lindsay}}, \bibinfo {author}
  {\bibfnamefont {D.}~\bibnamefont {Cherns}}, \bibinfo {author} {\bibfnamefont
  {J.~W.}\ \bibnamefont {Pomeroy}}, \bibinfo {author} {\bibfnamefont
  {S.}~\bibnamefont {Liu}}, \bibinfo {author} {\bibfnamefont {J.~H.}\
  \bibnamefont {Edgar}}, \ and\ \bibinfo {author} {\bibfnamefont
  {M.}~\bibnamefont {Kuball}},\ }\href {\doibase 10.1038/s42005-019-0145-5}
  {\bibfield  {journal} {\bibinfo  {journal} {Communications Physics}\ }\textbf
  {\bibinfo {volume} {2}} (\bibinfo {year} {2019}),\
  10.1038/s42005-019-0145-5}\BibitemShut {NoStop}%
\bibitem [{\citenamefont {Chng}\ \emph {et~al.}(2020)\citenamefont {Chng},
  \citenamefont {Zhu}, \citenamefont {Du}, \citenamefont {Wang}, \citenamefont
  {Whiteside}, \citenamefont {Ng}, \citenamefont {Shakerzadeh}, \citenamefont
  {Tsang},\ and\ \citenamefont {Teo}}]{Chng2020}%
  \BibitemOpen
  \bibfield  {author} {\bibinfo {author} {\bibfnamefont {S.~S.}\ \bibnamefont
  {Chng}}, \bibinfo {author} {\bibfnamefont {M.}~\bibnamefont {Zhu}}, \bibinfo
  {author} {\bibfnamefont {Z.}~\bibnamefont {Du}}, \bibinfo {author}
  {\bibfnamefont {X.}~\bibnamefont {Wang}}, \bibinfo {author} {\bibfnamefont
  {M.}~\bibnamefont {Whiteside}}, \bibinfo {author} {\bibfnamefont {Z.~K.}\
  \bibnamefont {Ng}}, \bibinfo {author} {\bibfnamefont {M.}~\bibnamefont
  {Shakerzadeh}}, \bibinfo {author} {\bibfnamefont {S.~H.}\ \bibnamefont
  {Tsang}}, \ and\ \bibinfo {author} {\bibfnamefont {E.~H.~T.}\ \bibnamefont
  {Teo}},\ }\href {\doibase 10.1039/D0TC02253E} {\bibfield  {journal} {\bibinfo
   {journal} {J. Mater. Chem. C}\ }\textbf {\bibinfo {volume} {8}},\ \bibinfo
  {pages} {9558} (\bibinfo {year} {2020})}\BibitemShut {NoStop}%
\bibitem [{\citenamefont {Giles}\ \emph {et~al.}(2018)\citenamefont {Giles},
  \citenamefont {Dai}, \citenamefont {Vurgaftman}, \citenamefont {Hoffman},
  \citenamefont {Liu}, \citenamefont {Lindsay}, \citenamefont {Ellis},
  \citenamefont {Assefa}, \citenamefont {Chatzakis}, \citenamefont {Reinecke},
  \citenamefont {Tischler}, \citenamefont {Fogler}, \citenamefont {Edgar},
  \citenamefont {Basov},\ and\ \citenamefont {Caldwell}}]{Giles2018-Natmat}%
  \BibitemOpen
  \bibfield  {author} {\bibinfo {author} {\bibfnamefont {A.~J.}\ \bibnamefont
  {Giles}}, \bibinfo {author} {\bibfnamefont {S.}~\bibnamefont {Dai}}, \bibinfo
  {author} {\bibfnamefont {I.}~\bibnamefont {Vurgaftman}}, \bibinfo {author}
  {\bibfnamefont {T.}~\bibnamefont {Hoffman}}, \bibinfo {author} {\bibfnamefont
  {S.}~\bibnamefont {Liu}}, \bibinfo {author} {\bibfnamefont {L.}~\bibnamefont
  {Lindsay}}, \bibinfo {author} {\bibfnamefont {C.~T.}\ \bibnamefont {Ellis}},
  \bibinfo {author} {\bibfnamefont {N.}~\bibnamefont {Assefa}}, \bibinfo
  {author} {\bibfnamefont {I.}~\bibnamefont {Chatzakis}}, \bibinfo {author}
  {\bibfnamefont {T.~L.}\ \bibnamefont {Reinecke}}, \bibinfo {author}
  {\bibfnamefont {J.~G.}\ \bibnamefont {Tischler}}, \bibinfo {author}
  {\bibfnamefont {M.~M.}\ \bibnamefont {Fogler}}, \bibinfo {author}
  {\bibfnamefont {J.~H.}\ \bibnamefont {Edgar}}, \bibinfo {author}
  {\bibfnamefont {D.~N.}\ \bibnamefont {Basov}}, \ and\ \bibinfo {author}
  {\bibfnamefont {J.~D.}\ \bibnamefont {Caldwell}},\ }\href {\doibase
  10.1038/nmat5047} {\bibfield  {journal} {\bibinfo  {journal} {Nature
  Materials}\ }\textbf {\bibinfo {volume} {17}},\ \bibinfo {pages} {134}
  (\bibinfo {year} {2018})}\BibitemShut {NoStop}%
\end{thebibliography}%


\end{document}